\documentclass[aps,prd,onecolumn,showpacs,amsmath,amssymb,nofootinbib]{revtex4}
\usepackage{amsfonts}
\usepackage{amsmath}
\usepackage{graphicx}
\usepackage{subfigure}
\usepackage{dcolumn}
\usepackage{bm}
\usepackage{booktabs}
\usepackage[utf8]{inputenc}
\usepackage{multirow}
\usepackage{graphicx,graphics,dcolumn,booktabs,bm}
\usepackage{longtable,lscape}
\usepackage{txfonts}
\usepackage{overpic}
\usepackage{amssymb}
\usepackage{indentfirst}
\usepackage{epsfig}
\usepackage{feynmf}   %{feynmp}
\usepackage{epstopdf}   %{feynmp}
\usepackage{slashed}  %for Feynman symbols
\usepackage{color}
\usepackage[section]{placeins}

\usepackage[colorlinks, citecolor=blue,anchorcolor=red,menucolor=red, linkcolor=red,filecolor=red,runcolor=red,urlcolor=blue,frenchlinks=red]{hyperref}

\usepackage{ulem}

\makeatletter
\newcommand{\figcaption}{\def\@captype{figure}\caption}
\newcommand{\tabcaption}{\def\@captype{table}\caption}

\newcommand{\Rmnum}[1]{\expandafter\@slowromancap\romannumeral #1@}
\def\hlinewd#1{%
  \noalign{\ifnum0=`}\fi\hrule \@height #1 \futurelet
   \reserved@a\@xhline}
\makeatother

\def\GG{\langle \alpha_{s}G^2\rangle}
\def\qGq{\langle \bar qGq \rangle}
\def\sGs{\langle \bar sGs \rangle}
\def\qq{\langle \bar qq \rangle }
\def\ss{\langle \bar ss \rangle }
\def\ms{m_s}
\def\bea{\begin{eqnarray}}
\def\eea{\end{eqnarray}}
\def\bq{\begin{quote}}
\def\eq{\end{quote}}

\def\ga{\left(}
\def\dr{\right)}

\def\non{\\ \nonumber}
														
\def\nnb{\nonumber}
\def\la{\langle}
\def\ra{\rangle}
\begin{document}
\title{Exotic tetraquark states with $J^{PC}=0^{+-}$}
\author{Yi-Chao Fu$^1$}
\email{fuych3@mail2.sysu.edu.cn}
\author{Zhuo-Ran Huang$^2$}
\email{huangzr@ihep.ac.cn}
\author{Zhu-Feng Zhang$^3$}
\email{zhangzhufeng@nbu.edu.cn}
\author{Wei Chen$^1$}
\email{chenwei29@mail.sysu.edu.cn}
\affiliation {$^1$School of Physics, Sun Yat-Sen University, Guangzhou 510275, China \\
$^2$Institute of High Energy Physics, Chinese Academy of Sciences, Beijing 100049, China \\
$^3$Physics Department, Ningbo University, Ningbo, 315211, China}

\begin{abstract}
We use the Laplace/Borel sum rules (LSR) and the finite energy/local duality sum rules (FESR) to investigate the non-strange $ud\bar u\bar d$ and hidden-strange $us\bar u\bar s$ tetraquark states with exotic quantum numbers $J^{PC}=0^{+-}$ . We systematically construct all eight possible tetraquark currents in this channel without covariant derivative operator. Our analyses show that the $ud\bar u\bar d$ systems have good behaviour of sum rule stability and expansion series convergence in both the LSR and FESR analyses, while the LSR for the $us\bar u\bar s$ states do not associate with convergent OPE series in the stability regions and only the FESR can provide valid results. We give the mass predictions $1.43\pm0.09$ GeV and $1.54\pm0.12$ GeV for the $ud\bar u\bar d$ and $us\bar u\bar s$ tetraquark states, respectively. Our results indicate that the $0^{+-}$ isovector $us\bar u\bar s$ tetraquark may only decay via weak interaction mechanism, e.g. $X_{us\bar{u}\bar{s}}\to K\pi\pi$, since its strong decays are forbidden by kinematics and the symmetry constraints on the exotic quantum numbers. It is predicted to be very narrow, if it does exist. The $0^{+-}$ isoscalar $us\bar u\bar s$ tetraquark is also predicted to be not very wide because its dominate decay mode $X_{us\bar{u}\bar{s}}\to\phi\pi\pi$ is in  
$P$-wave.
\end{abstract}

\maketitle
\section{Introduction}
In the constituent quark model, mesons consist of a pair of quark and antiquark ($q\bar q$)~\cite{2007-Klempt-p1-202, PhysRevD.98.030001}. They can be
characterized by the isospin $I$, the total angular momentum $J$, the parity $P$, and the charge-conjugation parity $C$ (for charge neutral states). For a fermion-antifermion system ($q\bar q$), these quantum numbers are given as
\begin{eqnarray}
I=0, 1, \, J=0, 1, 2\cdots, \, P=(-1)^{L+1}\, , \, C=(-1)^{L+S}\, ,
\end{eqnarray}
where $L$ is the relative orbit angular momentum between $q$ and $\bar q$ and $S$ the total spin. For charged mesons, it is useful to define the $G$ parity instead of $C$ parity
\begin{eqnarray}
G=(-1)^IC=(-1)^{L+S+I}\, .
\end{eqnarray}
For such meson states, the allowed $J\leq2$ quantum numbers are $J^{PC}=0^{-+}\, , 0^{++}\, , 1^{--}\, , 1^{+-}\, , 1^{++}\, , 2^{--}\, , 2^{-+}\, , 2^{++}$. The combinations $J^{PC}=0^{--}\, , 0^{+-}\, , 1^{-+}\, , 2^{+-}$
are not allowed for the conventional $q\bar q$ systems. In other words, they are exotic quantum numbers in the quark model.

However, these exotic quantum numbers can be reached in other configurations such as hybrids~\cite{Ho:2018cat,Zhang:2013rya,Huang:2014hya,Huang:2016upt,Chetyrkin:2000tj}, glueballs~\cite{Qiao:2015iea,Pimikov:2017bkk} and tetraquarks, which are not forbidden by QCD itself. Tetraquarks ($qq\bar q\bar q$) are bound states of diquarks ($qq$) and antidiquarks ($\bar q\bar q$). Their existence  was firstly suggested by R.  Jaffe in 1977~\cite{1977-Jaffe-p267-267,1977-Jaffe-p281-281}. The light scalar mesons have been considered as good candidates of tetraquarks \cite{2004-Maiani-p212002-212002,2007-Chen-p94025-94025,2007-Chen-p369-372}. For hadrons with exotic quantum numbers, the $1^{-+}$ hybrid meson has been extensively studied since it was predicted to be the
lightest hybrid state~\cite{2015-Meyer-p21-58}. Especially, there are now some evidence on the existence of such hybrid mesons~\cite{1997-Thompson-p1630-1633,1999-Abele-p349-355,2007-Adams-p27-31}. The $1^{-+}$ light tetraquarks have also been studied in Refs.~\cite{2008-Chen-p117502-117502,2008-Chen-p54017-54017} to predict their masses for both $I=0$ and $1$ channels. In Refs.~\cite{2009-Jiao-p114034-114034,2017-Huang-p76017-76017}, the light tetraquark state with $J^{PC}=0^{--}$ was predicted to be the possible $\rho\pi$ dominance in the $D^0$ decay.

To date, the studies for other exotic quantum numbers $J^{PC}=0^{+-}\, , 2^{+-}$ are much less appealing. In Ref.~\cite{Du:2012pn}, the tetraquark states with $J^{PC}=0^{+-}$ were studied for both light and heavy sectors in QCD sum rules. The authors used the tetraquark currents containing a covariant derivative, which will increase the dimension of the interpolating operators. Finally, the light $0^{+-}$ tetraquark was concluded not exist due to the bad OPE series behavior. In this paper, we shall revisit the light $0^{+-}$ tetraquarks by using the interpolating currents without derivative. These currents have lower dimension than those in Ref.~\cite{Du:2012pn}, which will result in better OPE behaviors and mass predictions. We shall use both the Laplace sum rules (LSR) and finite energy sum rules (FESR) to perform numerical analyses. The possible decay patterns of the $0^{+-}$ tetraquark states will be discussed at last.
\section{Laplace sum rules and finite energy sum rules}
In this section, we introduce the formalism of QCD sum rules, which has been a very useful method to study the hadronic properties in the past several decades~\cite{1979-Shifman-p385-447,1985-Reinders-p1-1,2000-Colangelo-p1495-1576}. For a vector current of the general form $j_{\mu}(x)$, we consider the two-point correlation function
\bea
\nonumber
\Pi_{\mu\nu}(q^{2}) & = & i\int d^{4}xe^{iqx}\left\langle 0\left|T\left[j_{\mu}(x)j_{\nu}^{+}(0)\right]\right|0\right\rangle \label{eq:1}\\
 & = & (q_{\mu}q_{\nu}-q^{2}g_{\mu\nu})\Pi_{v}(q^{2})+q_{\mu}q_{\nu}\Pi_{s}(q^{2})\, ,
\eea
where $\Pi_{v}(q^{2})$ and $\Pi_{s}(q^{2})$ are the invariant functions receiving the contributions from the corresponding pure vector and scalar intermediate states, respectively. The invariant functions obey the dispersion relation, which relates the $\Pi(q^2)$ with the spectral function
\begin{eqnarray}
\Pi(q^{2})=\left(q^{2}\right)^{N}\int_{0}^{\infty}\mathrm{d}s\frac{\rho\left(s\right)}{s^{N}\left(s-q^{2}-\mathrm{i}\epsilon\right)}+\sum_{k=0}^{N-1}b_{n}\left(q^{2}\right)^{k}\, ,
\label{eq:20}
\end{eqnarray}
where $b_n$ are the unknown subtraction constants and they can be removed by performing the Borel transformation to $\Pi(q^2)$. On the QCD side, we can perturbatively calculate the correlation function by using the operator product expansion (OPE) method. In such calculations, the correlation function $\Pi(q^2)$ is expressed as power expansion series using the QCD vacuum condensates of increasing dimensions.

On the hadronic side, the spectral functions are expressed via the quark hadron duality
\bea
\nonumber
\rho(s)\equiv\frac{1}{\pi}\textrm{Im}\Pi_{v/s}(s)&\simeq&\sum_n\delta(s-m_n^2)\langle0|\eta|n\rangle\langle n|\eta^+|0\rangle\\
&\simeq&f_H^2\delta(s-m_H^2)+ \cdots\, , \label{eq:rho}
\eea
where we use the ``one single narrow resonance ansatz"~\cite{Narison:1980ti,Narison:2002hk}. ``$\cdots$" denotes the contributions of the higher excited states and the QCD continuum, and $m_H$ and $f_H$ are the mass and the coupling of the lowest lying hadron state. One should note that we omit the tensor structure in the second step if $\eta$ be a vector current. Under this duality ansatz, one can match the QCD side of the correlation function with the hadronic side, and then obtain the sum rules for the hadron parameters such as the hadron mass, the coupling constant and the magnetic moment. Technically, one usually applies the Borel transformation to $\Pi(q^2)$ at both sides in order to enhance the contribution of the lowest-lying state and improve the OPE convergence. Finally, the Laplace/Borel sum rules (LSR) moment can be derived as
\bea\label{usr}
{M}(\tau,s_0)
&=& \int_{0}^{s_0} {ds}~e^{-s\tau}\rho(s)\, .
\eea
Inserting the spectral function in Eq.~\eqref{eq:rho}, we extract the lowest-lying hadron mass as the following ratio
\bea\label{usr}
{R}(\tau,s_0)&=& -\frac{d}{d\tau} \ln {
M}=M_H^2\, ,
\eea
in which $\tau$ is the Borel parameter and $s_0$ is the continuum threshold above which the contributions from the higher excited states and QCD continuum can be approximated well by the spectral function. It is clear that the
hadron mass $M_H$ will depend on these two parameters $(\tau, s_0)$. To establish reliable sum rule analyses, one needs to pick out suitable $(\tau, s_0)$ working range to make sure the validity of the OPE truncation and the suppression of the continuum contribution. In this parameter working range, we would expect that the mass curves are insensitive to the variation of $\tau$ and $s_0$, which will finally provide reliable predictions on the hadron masses.

As is well-known, the LSR for light tetraquarks usually suffer from higher dimension of the interpolating currents which lead to poor OPE convergence or absence of the sum rule stability. The finite energy sum rule, also known as the local duality sum rule, can be a valid complementary method. FESR can be obtained either by applying the Cauchy theorem to the correlator on a contour with radius $r=s_0$~\cite{Shankar:1977ap} or by simply letting the Borel parameter $\tau$ vanish in LSR. This method reduces the effects from the power corrections to the sum rule moment and has been shown to be useful in studying multi-quark states~\cite{Narison:2009vj,2017-Huang-p76017-76017,2008-Chen-p54017-54017,Matheus:2006xi}\footnote{For very recent applications of FESR in studying the decay constants of heavy-light mesons, see~\cite{Lucha:2018ouu,Lucha:2017zng}}. The nth moment and ratio of FESR read
\begin{eqnarray}
W(n\, ,s_0)&=& \int^{s_0}_0 \rho(s)s^n ds\, , \label{moment}
\\
M_H^2(n\, ,s_0)&=&\frac{W(n+1\, ,s_0)}{W(n\, ,s_0)}.
\end{eqnarray}
%\bea\label{fesr}
%{\cal R}_{v/s}(s_0)
%&=&{\int_{0}^{s_0} {ds}~s
%~\frac{1}{\pi}~\mbox{Im} \Pi_{v/s}(s)\over
%\int_{0}^{s_0} {ds}
%~\frac{1}{\pi}~\mbox{Im} \Pi_{v/s}(s)} \simeq M_H^2~,
%\eea
In this work we use the 0th moment ratio which enhances the convergence of the FESR expansion series in Eq.~\eqref{moment}.

%========================================
\section{Tetraquark interpolating currents with $J^{PC}=0^{+-}$}
%========================================
In this section, we construct the tetraquark interpolating currents with exotic quantum numbers $J^{PC}=0^{+-}$.
In Ref.~\cite{Du:2012pn}, the exotic $0^{+-}$ light tetraquarks have been studied by using the tetraquark currents containing covariant derivatives. However, their calculations do not support the existence of such tetraquarks since there is no stable mass sum rule there. Such bad sum rule behavior appears due to the special Lorentz structures
of the interpolating currents adopted in Ref.~\cite{Du:2012pn}. The covariant derivative draw a $P$-wave excitation to the current, which will result in unstable mass sum rules. In this work, we shall revisit the light $0^{+-}$ tetraquarks by constructing the interpolating currents without derivative. These lower dimension currents may lead to better OPE behavior and mass prediction.

We compose the light tetraquark currents by using six distinct diquark operators in Lorentz space:  $q^T_aCq_b$, $q^T_a C\gamma_5q_b$, $q^T_aC\gamma_\mu q_b$, $q^T_aC\gamma_\mu\gamma_5q_b$, $q^T_a C\sigma_{\mu\nu}q_b$, $q^T_aC\sigma_{\mu\nu}\gamma_5q_b$. The diquark-antidiquark type of tetraquark currents without derivatives are then constructed as
\begin{eqnarray}
\nonumber J_{1\mu}&=&u^T_aC\gamma_5d_b(\bar{u}_a\gamma_{\mu}\gamma_5C\bar{d}^T_b+\bar{u}_b\gamma_{\mu}\gamma_5C\bar{d}^T_a)
-
u^T_aC\gamma_{\mu}\gamma_5d_b(\bar{u}_a\gamma_5C\bar{d}^T_b+\bar{u}_b\gamma_5C\bar{d}^T_a)\,
, \non J_{2\mu}&=&u^T_aC\gamma^{\nu}d_b(\bar{u}_a\sigma_{\mu\nu}C\bar{d}^T_b-\bar{u}_b\sigma_{\mu\nu}C\bar{d}^T_a)
-
u^T_aC\sigma_{\mu\nu}d_b(\bar{u}_a\gamma^{\nu}C\bar{d}^T_b-\bar{u}_b\gamma^{\nu}C\bar{d}^T_a)\,
, \non J_{3\mu}&=&u^T_aC\gamma_5d_b(\bar{u}_a\gamma_{\mu}\gamma_5C\bar{d}^T_b-\bar{u}_b\gamma_{\mu}\gamma_5C\bar{d}^T_a)
-
u^T_aC\gamma_{\mu}\gamma_5d_b(\bar{u}_a\gamma_5C\bar{d}^T_b-\bar{u}_b\gamma_5C\bar{d}^T_a)\,
,
\\ \label{currents2}
J_{4\mu}&=&u^T_aC\gamma^{\nu}d_b(\bar{u}_a\sigma_{\mu\nu}C\bar{d}^T_b+\bar{u}_b\sigma_{\mu\nu}C\bar{d}^T_a)
-
u^T_aC\sigma_{\mu\nu}d_b(\bar{u}_a\gamma^{\nu}C\bar{d}^T_b+\bar{u}_b\gamma^{\nu}C\bar{d}^T_a)\,
, \non J_{5\mu}&=&u^T_aCd_b(\bar{u}_a\gamma_{\mu}C\bar{d}^T_b+\bar{u}_b\gamma_{\mu}C\bar{d}^T_a)
-
u^T_aC\gamma_{\mu}d_b(\bar{u}_aC\bar{d}^T_b+\bar{u}_bC\bar{d}^T_a)\,
, \non J_{6\mu}&=&u^T_aC\gamma^{\nu}\gamma_5d_b(\bar{u}_a\sigma_{\mu\nu}\gamma_5C\bar{d}^T_b+\bar{u}_b\sigma_{\mu\nu}\gamma_5C\bar{d}^T_a)
-
u^T_aC\sigma_{\mu\nu}\gamma_5d_b(\bar{u}_a\gamma^{\nu}C\bar{d}^T_b+\bar{u}_b\gamma^{\nu}C\bar{d}^T_a)\,
, \non J_{7\mu}&=&u^T_aCd_b(\bar{u}_a\gamma_{\mu}C\bar{d}^T_b-\bar{u}_b\gamma_{\mu}C\bar{d}^T_a)
-
u^T_aC\gamma_{\mu}d_b(\bar{u}_aC\bar{d}^T_b-\bar{u}_bC\bar{d}^T_a)\,
, \non J_{8\mu}&=&u^T_aC\gamma^{\nu}\gamma_5d_b(\bar{u}_a\sigma_{\mu\nu}\gamma_5C\bar{d}^T_b-\bar{u}_b\sigma_{\mu\nu}\gamma_5C\bar{d}^T_a)
-
u^T_aC\sigma_{\mu\nu}\gamma_5d_b(\bar{u}_a\gamma^{\nu}\gamma_5C\bar{d}^T_b-\bar{u}_b\gamma^{\nu}\gamma_5C\bar{d}^T_a)\, ,
\end{eqnarray}
where $a, b$ are color indices, $T$ is the transposition operator and $C$ the charge conjugation operator. These interpolating currents in Eq.~\eqref{currents2} can couple to both the $J^{PC}=0^{+-}$ and $1^{--}$ channels, which will induce the scalar $\Pi_s(q^2)$ and vector $\Pi_v(q^2)$ respectively in Eq.~\eqref{eq:1}. In this work, we focus on the exotic $0^{+-}$ channel in our calculation. By replacing $d\to s$ in Eq.~\eqref{currents2}, we obtain the corresponding $us\bar{u}\bar{s}$ tetraquark currents with $J^{PC}=0^{+-}$. Using these interpolating currents, we calculate their two-point correlation functions and the
spectral functions. We shall study both the $ud\bar{u}\bar{d}$ and $us\bar{u}\bar{s}$ tetraquark systems in the following section.
\section{QCD expressions for the two-point correlation functions}
Using the interpolating currents listed above, we obtain the QCD expressions for the corresponding two-point correlation functions via the standard technique of SVZ expansion \cite{Shifman:1984wx}. Up to dimension-8 condensate terms in the power expansion and leading-order contributions in the perturbative expansion, the general expression for the LSR moment corresponding to the $ud\bar{u}\bar{d}$-type and the $us\bar{u}\bar{s}$-type currents respectively read
\begin{flalign}
\label{eq:udud}
M^{ud\bar{u}\bar{d}}_{0}(\tau,s_0)=&\int^{s_0}_{0}\rho^{ud\bar{u}\bar{d}}_{0}(s)e^{-\tau s}ds=a_{i}\frac{e^{-s_0 \tau } \lbrace -s_0 \tau \lbrack s_0 \tau  (s_0 \tau +3)+6\rbrack -6\rbrace+6}{\tau ^4}+b_{i}\frac{1-e^{-s_0 \tau } (s_0 \tau +1)}{\tau ^2}{}\nonumber\\
&-c_{i}\frac{1-e^{-s_0 \tau }}{\tau }+ d_{i} \lbrack 2+\ln(4\pi)-\ln(\frac{1}{\tau\tilde\mu^2})+\Gamma(0,s_0 \tau)\rbrack\, ,
\end{flalign}
and
\begin{flalign}
\label{eq:usus}
M^{us\bar{u}\bar{s}}_{0}(\tau,s_0)=&\int^{s_0}_{0}\rho^{us\bar{u}\bar{s}}_{0}(s)e^{-\tau s}ds = -a^{\prime}_{i} \frac{e^{-s_0 \tau }\lbrace -s_0 \tau \lbrack s_0 \tau  (s_0 \tau +3)+6\rbrack -6\rbrace+6}{\tau ^4}-b^{\prime}_{i} \frac{e^{-s_0 \tau } \lbrack -s_0 \tau  (s_0 \tau +2)-2 \rbrack +2}{\tau ^3}\nonumber\\
&-c^{\prime}_{i} \frac{1-e^{-s_0 \tau } (s_0 \tau +1)}{\tau ^2}
- d^{\prime}_{i}\frac{1-e^{-s_0 \tau }}{\tau }+e^{\prime}_{i}+ f^{\prime}_{i}\lbrack \gamma_{E}-\ln(\frac{1}{\tau\tilde\mu^2})+\Gamma(0,s_0 \tau)\rbrack\, ,
\end{flalign}
where $\tilde\mu=1$ GeV, $\gamma_{E}$ the Euler constant, and $\Gamma$ the incomplete Gamma function. The values of the QCD condensates are listed in Table.~\ref{param_tab}, and $a_i$-$d_i$, $a^{\prime}_i$-$f^{\prime}_i$ are the Wilson coefficients. We list their expressions in APPENDIX~\ref{app:A}.
\begin{table}
\renewcommand\arraystretch{1.5}
\caption{\label{param_tab}QCD parameters used in our analysis: $\rho$ indicates the violation of factorization hypothesis.}
\begin{tabular}{cc}
  \hlinewd{.8pt}
   QCD parameters& Reference \\
   \hline
  $\la\alpha_s G^2\ra \simeq (7\pm2)\times10^{-2}~\rm{GeV}^4$ &  \cite{Launer:1983ib,Narison:1995jr,Eidelman:1978xy,Narison:2011xe,Narison:2011rn} \\
  $g\la\bar{\psi}G\psi\ra\equiv g\la\bar{\psi}\frac{\lambda_a}{2}\sigma^{\mu\nu}G^a_{\mu\nu}\psi\ra\simeq
(0.8\pm
0.1)~{\rm GeV}^2\la\bar \psi\psi\ra$ &  \cite{Dosch:1988vv,Ioffe:1981kw} \\
  $\rho\alpha_s  \la\bar \psi\psi\ra^2\simeq
(4.5\pm 0.3) \times 10^{-4}~\rm{ GeV}^6$ &  \cite{Narison:1992ru,Narison:1995jr,Narison:2009vy}  \\
  $\Lambda_{\rm QCD}= (353\pm 15)~{\rm MeV}$ &  \cite{Narison:2009vy} \\
  $\la \bar{s}s \ra / \la \bar{u}u \ra=0.74 \pm 0.03$ &\cite{NARISON2015189}\\
  $m_s=95^{+9}_{-3} ~\rm{MeV}$ &\cite{PhysRevD.98.030001}\\
  \hlinewd{.8pt}
\end{tabular}
\end{table}
\section{LSR and FESR numerical analyses}
In this section, we shall perform our numerical analyses for the non-strange and hidden-strange $0^{+-}$ tetraquark states using both the LSR and the FESR methods. The methodology adopted in this work directly follow from \cite{2017-Huang-p76017-76017} and the references therein, where one can find more details about the sum rule stability criteria applied.
\subsection{Analysis for the $ud\bar{u}\bar{d}$ tetraquark states}
 We first focus on the $ud\bar{u}\bar{d}$-type tetraquark systems. From the Wilson coefficients listed in APPENDIX~\ref{app:A}, we find that there exist degeneracies between the OPE results corresponding to the $ud\bar{u}\bar{d}$-type $J_1$/$J_5$, $J_2$/$J_8$, $J_3$/$J_7$ and $J_4$/$J_6$ in the chiral limit ($m_u=m_d=0$). Therefore, we will only present the analyses for $J_1$-$J_4$.

While applying the LSR analysis, we require both the $\tau$ and $s_0$ stability in order to get solid predictions. Practically, we read the predictions from the extremum points of the $m_H-\tau$ and $m_H-s_0$ curves, from which continuum threshold can also be rigorously determined. In the case where the LSR stability is reached, the OPE convergence at the given stability points should be further checked before taking into account the obtained mass values in the final estimation. For the $ud\bar{u}\bar{d}$ currents, we find that all the associated LSR moment ratios reach stability, but only those corresponding to $J_3$ ($J_7$) and $J_4$ ($J_6$) have converging OPE series. For $J_1$ ($J_5$) and $J_2$ ($J_8$), the highest order power corrections (dimension-8 condensate terms) contribute more than $10\%$ in the OPE series, rendering a problematic truncation. Therefore we only consider results obtained using $J_3$ ($J_7$) and $J_4$ ($J_6$) for the final mass determination.

\begin{figure}[htbp]
\centering
\subfigure[]{\label{fig:LSRududj3j41}
\includegraphics[scale=0.7]{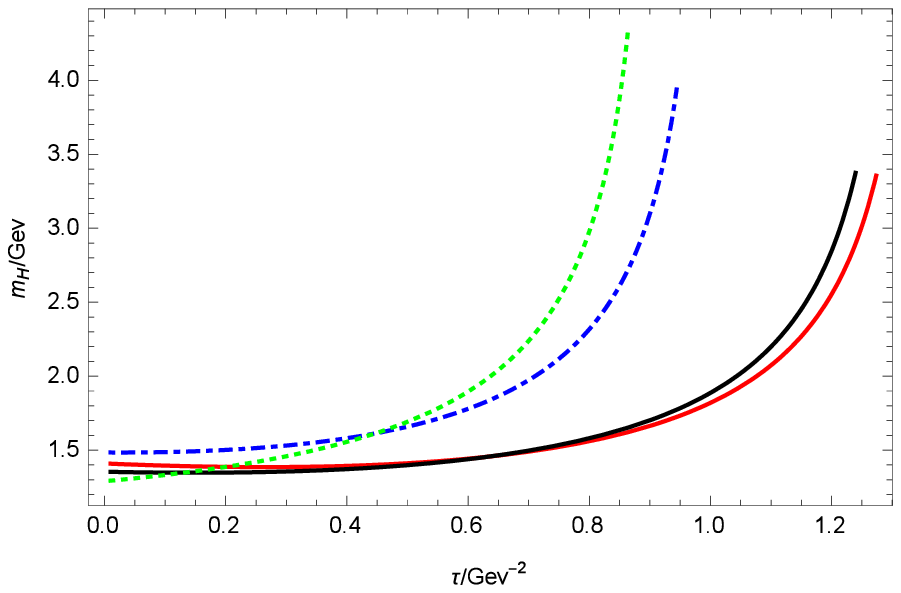}}
\subfigure[]{\label{fig:LSRududj3j42}
\includegraphics[scale=0.7]{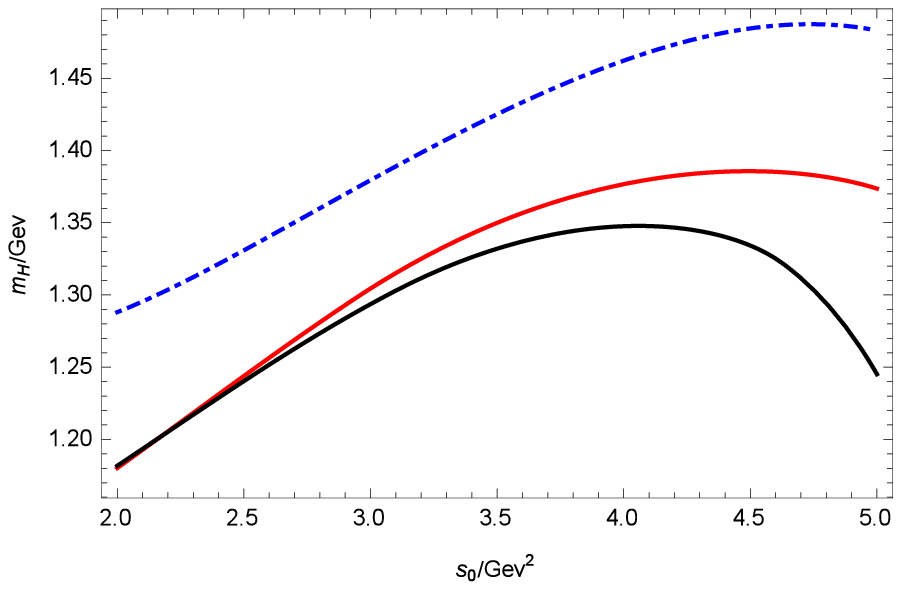}}
\caption{\label{ududlsr}(a) $ud\bar{u}\bar{d}$ four quark state mass versus $\tau$ in LSR obtained using $J_3(J_7)$ (red continuous), $J_3(J_7)$ considering violation of factorization (blue dotted-dashed), $J_4(J_6)$ (black continuous) and $J_4(J_6)$ considering violation of factorization (green dotted); (b) the same as (a) but for mass versus $s_0$.
}
\label{fig:LSRududj3j4}
\end{figure}

The behaviours of the LSR ratios corresponding to $J_3$ ($J_7$) and $J_4$ ($J_6$) are shown in FIG.~\ref{fig:LSRududj3j4}, where the left panel shows the $\tau$-stability (if any) and the right panel shows the $s_0$-stability. One can see that the LSR ratio corresponding to $J_3$ ($J_7$) and $J_4$ ($J_6$) reach the $\tau$ and $s_0$ stability when the values of the dimension-8 condensates are estimated using the vacuum saturation approximation. However, the mass curve corresponding to $J_4$ ($J_6$) becomes monotonous when factorization is violated by a factor of 2 (see the green curve in FIG.~\ref{fig:LSRududj3j41}). In contrast, the LSR curves  corresponding to $J_3$ ($J_7$) have both the $\tau$ and $s_0$ stability with and without considering the violation of factorization, which can provide an error estimation due to violation of factorization. From FIG.~\ref{fig:LSRududj3j42} we obtain the mass predictions at the $s_0$-stability points (extremums) as below
\begin{eqnarray}
M_{J_{3}/J_7;LSR}&=&1.39(1.49)~{\rm GeV}~at~s_0=4.50(4.75)~{\rm GeV^2}\, ,\nnb\\
M_{J_{4}/J_6;LSR}&=&1.35~{\rm GeV}~at~s_0=4.09~{\rm GeV^2}\, ,
\end{eqnarray}
 where the values in the parentheses are obtained by taking into account the violation of factorization of the dimension-8 condensates by a factor of 2. For $J_1$ ($J_5$) and $J_2$ ($J_8$), we obtain similar curves as those in FIG.~\ref{fig:LSRududj3j4}, but the stability points associate with poorly OPE convergence therefore we do not consider these values. In TABLE~\ref{tab:ududlsrconvergence}, we present the behaviour of OPE at the stability points of $\tau$ and $s_0$.
%%%%%%%%%%%%%%%%%%%%%%%%%%%%%%%1
%%%%%%%%%%%%%%%%%%%%%%%%%%%%%%%%%%%%
\begin{table}[htbp]
\renewcommand\arraystretch{1.5}
\centering
\caption{\label{tab:ududlsrconvergence}OPE terms at the LSR stability points using the $ud\bar{u}\bar{d}$ currents.}
\begin{tabular}{cccccccc}
  \hlinewd{.8pt}
    $J_i$ &${1\over\tau}{\hat B}\Pi_{i}^{d=0}/{\rm GeV}^4$ & ${1\over\tau}{\hat B}\Pi_{i}^{d=4}/{\rm GeV}^4$ & ${1\over\tau}{\hat B}\Pi_{i}^{d=6}/{\rm GeV}^4$ & ${1\over\tau}{\hat B}\Pi_{i}^{d=8}/{\rm GeV}^4$ & ${1\over\tau}{\hat B}\Pi_{i}^{d=8}$/OPE & $\tau$/${\rm GeV}^{-2}$ & $s_0/{\rm GeV}^2$\\
    \hline
   $J_1/J_5$ & $5.77935\times 10^{-6}$& $-7.94295\times 10^{-7}$& $-8.64314\times 10^{-6}$&$-9.16139\times 10^{-7}$ &$0.200283$ &$0.433$&5.33 \\
   $J_2/J_8$ &$2.12627\times 10^{-5}$ &$1.24396\times 10^{-6}$ & $-1.62246\times10^{-5}$&$-2.38152\times10^{-6}$ &-0.610577 &0.346 &4.89 \\
   $J_3/J_7$ &$2.05976\times 10^{-5}$ &$2.12063\times10^{-6}$ &$-7.06129\times10^{-6}$ &$-1.1931\times10^{-6}$ &-0.0824886 &0.265 &4.50 \\
   $J_4/J_6$ &$1.2703\times10^{-3}$ &$7.47871\times10^{-5}$ &$-7.58611\times10^{-5}$ &$-1.23908\times10^{-5}$ &-0.00985874 &0.148 &4.09 \\
  \hlinewd{.8pt}
\end{tabular}
\end{table}

\begin{figure}[htbp]
\renewcommand\arraystretch{1.5}
\centering
\subfigure[]{\label{fig:ududfesr1}
\includegraphics[scale=0.7]{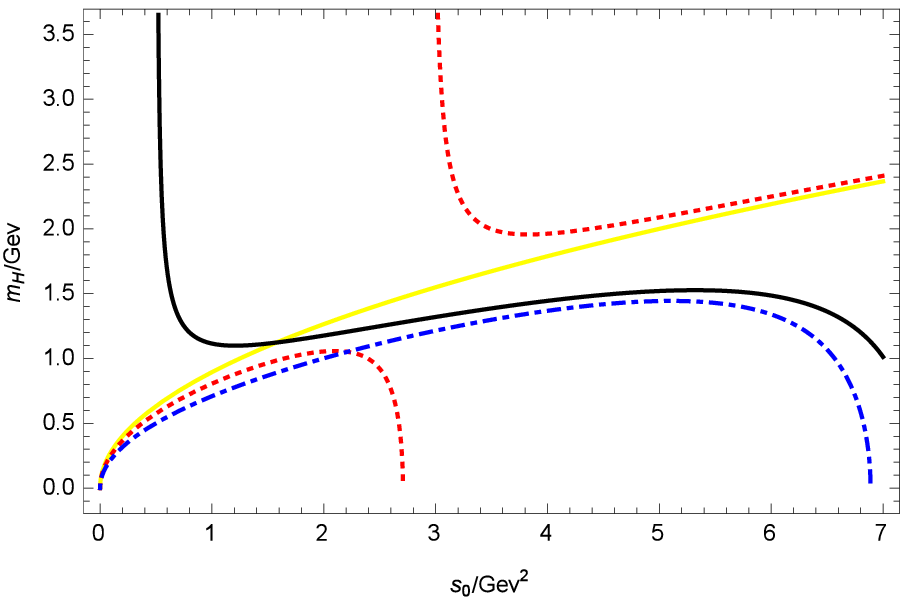}}
\subfigure[]{\label{fig:ududfesr2}
\includegraphics[scale=0.7]{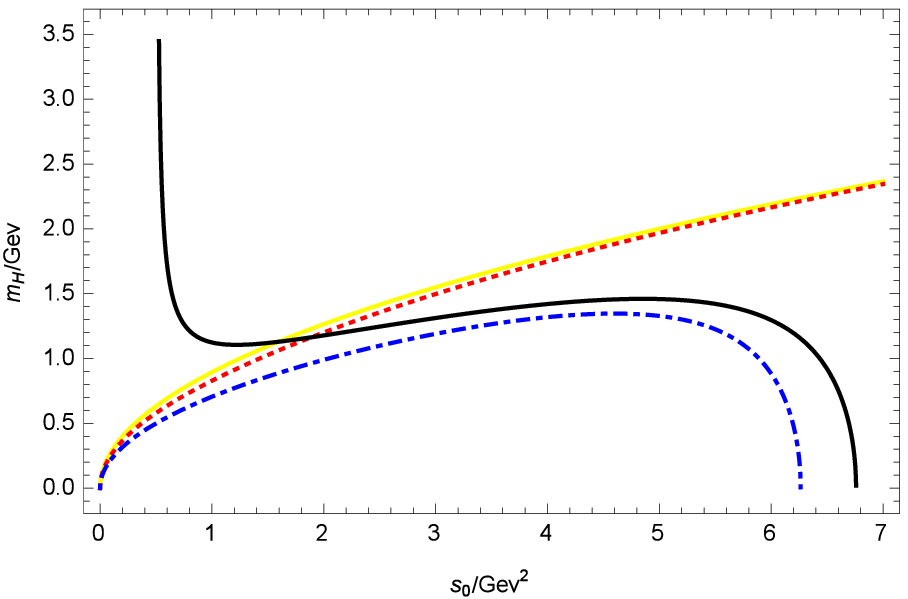}}\\
\subfigure[]{\label{fig:ududfesr3}
\includegraphics[scale=0.7]{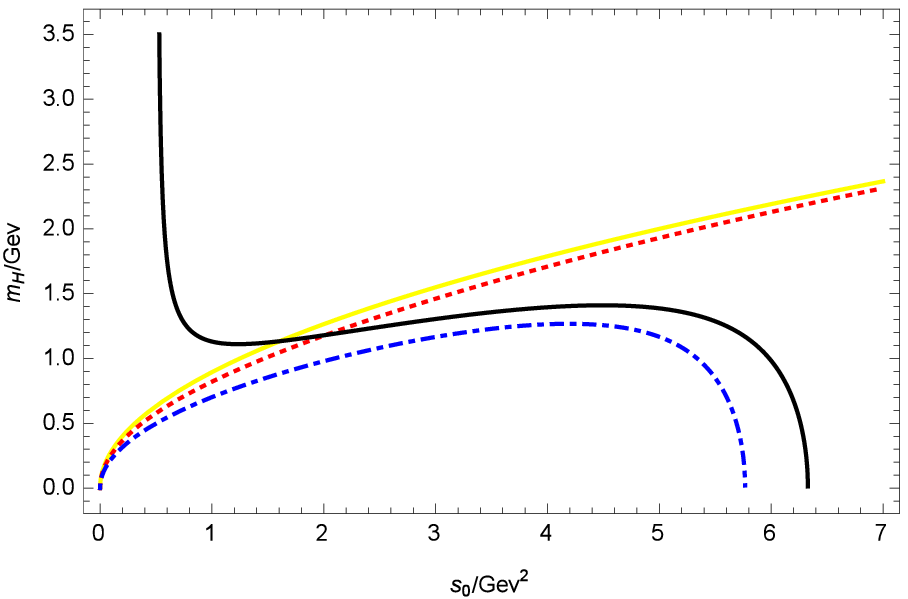}}
\subfigure[]{\label{fig:ududfesr4}
\includegraphics[scale=0.7]{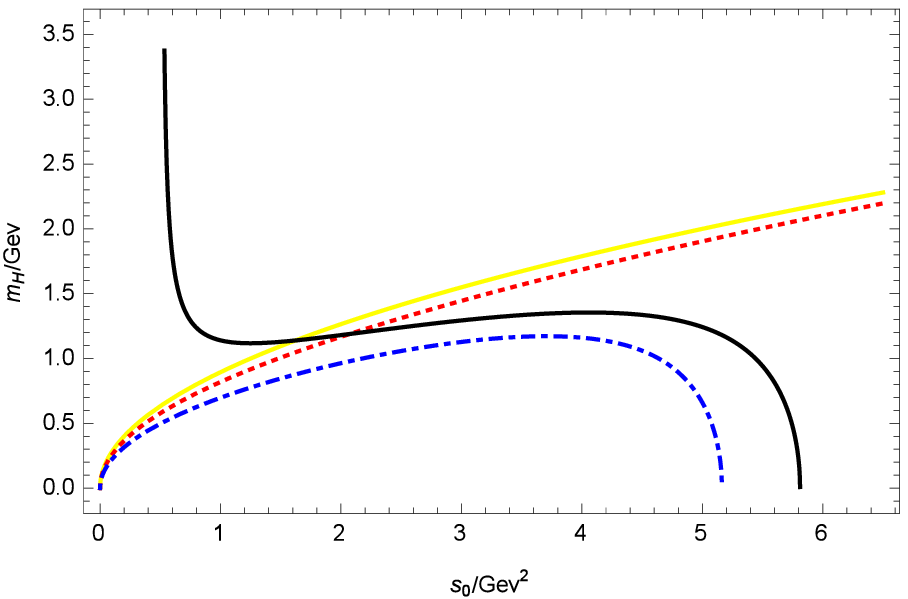}}\\

\caption{\label{fig:ududfesr}(a) $ud\bar{u}\bar{d}$ four quark state mass versus $s_0$ in FESR obtained using $J_1(J_5)$ considering in the OPE series the perturbative terms (yellow continuous), d=4 condensate terms (red dotted), d=6 condensate terms (blue dotted-dashed), d=8 condensate terms (black continuous); (b), (c), (d) are the same as (a) but for $J_2(J_8)$, $J_3(J_7)$, $J_4(J_6)$ respectively.
}
\end{figure}

In FIG.~\ref{fig:ududfesr}, we show the FESR curves obtained by truncating the OPE at different orders. One can see that for all currents the corresponding FESR moment ratios increase gradually for considering only the perturbative terms. With inclusion of the dimension-4 or dimension-6 condensate contributions, the mass curves start to present inflexions or stability points. One can then read the optimal mass values from these stability points. As shown in FIG.~\ref{fig:ududfesr}, the stability points of the FESR curves obtained by considering the condensate contributions up to dimension-8 (black continuous curves) are close to those of the curves obtained by only considering $d\leqq6$ condensate terms (blue dotted-dashed curves), except for the curve corresponding to $J_4$ ($J_6$). The difference is about 13\% for $J_4$ ($J_6$) while less than 10\% for other currents, which indicates that the OPE truncation for $J_4$ ($J_6$) may associate with relatively large uncertainties. Accordingly, we retain the mass predictions in FESR as
%%%%%%%%%%%%%%%%%%%%%%%%%%%%%%%1
%%%%%%%%%%%%%%%%%%%%%%%%%%%%%%%%%%%%
\begin{eqnarray}
M_{J_{1}/J_5;FESR}&=&1.53(1.57)~{\rm GeV}~at~s_0=5.32(5.43)~{\rm GeV^2},\nnb\\
M_{J_{2}/J_8;FESR}&=&1.46(1.49)~{\rm GeV}~at~s_0=4.86(4.85)~{\rm GeV^2},\nnb\\
M_{J_{3}/J_7;FESR}&=&1.41(1.44)~{\rm GeV}~at~s_0=4.49(4.41)~{\rm GeV^2}.
\end{eqnarray}
Taking the arithmetic average of the valid LSR and FESR results, we give our prediction for the $0^{+-}$ $ud\bar{u}\bar{d}$ tetraquark mass as
\begin{eqnarray}
M_{ud\bar{u}\bar{d}}&=&1.43\pm0.09~{\rm GeV}\, ,
\end{eqnarray}
where the error comes from the uncertainties of the QCD parameters, the results from different interpolating currents and the violation of factorization.
\subsection{Analysis for the $us\bar{u}\bar{s}$ tetraquark states}
\begin{figure}[htbp]
\centering
\subfigure[]{
\includegraphics[scale=0.7]{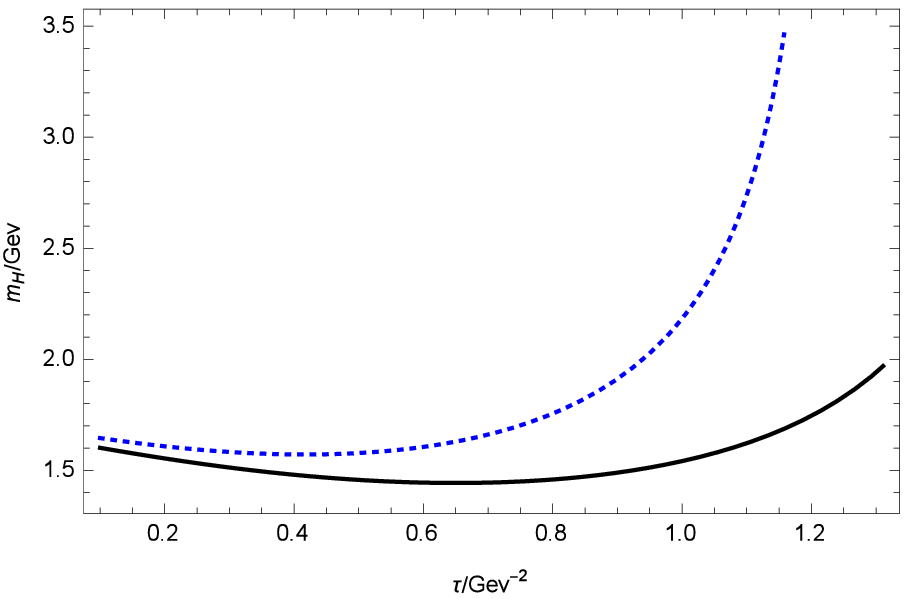}}
\subfigure[]{
\includegraphics[scale=0.7]{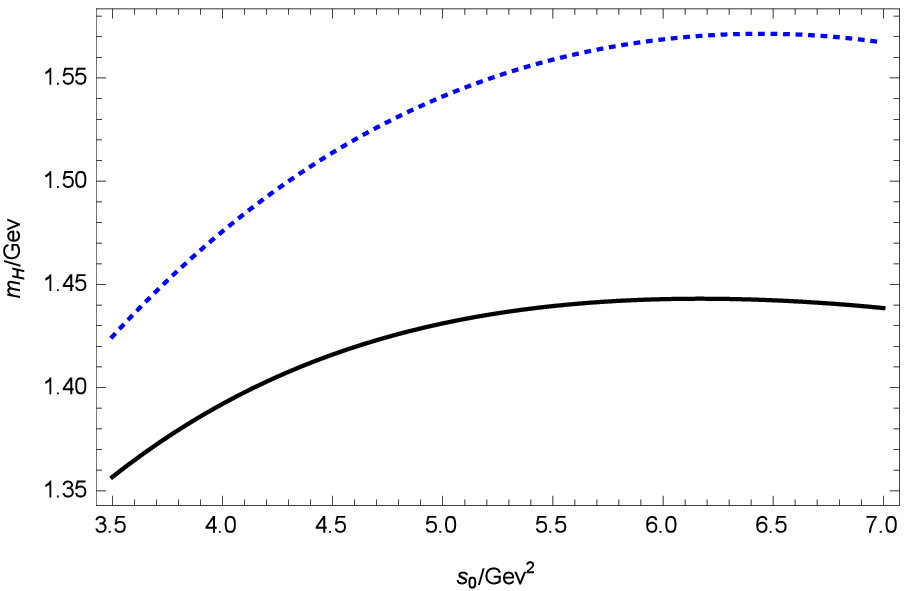}}\\
\caption{(a) $us\bar{u}\bar{s}$ four quark state mass versus $\tau$ in LSR obtained using $J_1$ (black continuous) and $J_1$ considering violation of factorization(blue dotted-dashed); (b) the same (a) but for mass versus $s_0$.
}
\label{fig:ususlsrj3}
\end{figure}

\begin{figure}[htbp]
\centering
%\subfigure[]{
\includegraphics[scale=0.8]{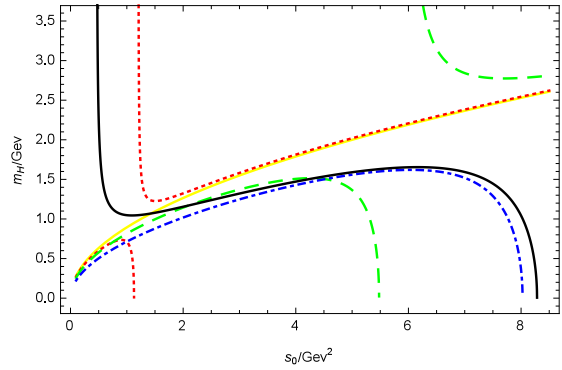}
%}
%\subfigure[]{
%\includegraphics[scale=0.55]{0+-ususfesrj2convergence.pdf}}\\
%\subfigure[]{
%\includegraphics[scale=0.55]{0+-ususfesrj3convergence.pdf}}
%\subfigure[]{
%\includegraphics[scale=0.55]{0+-ususfesrj4convergence.pdf}}\\
%\subfigure[]{
%\includegraphics[scale=0.55]{0+-ususfesrj5convergence.pdf}}
%\subfigure[]{
%\includegraphics[scale=0.565]{0+-ususfesrj6convergence.pdf}}\\
%\subfigure[]{
%\includegraphics[scale=0.55]{0+-ususfesrj7convergence.pdf}}
%\subfigure[]{
%\includegraphics[scale=0.55]{0+-ususfesrj8convergence.pdf}}\\
\caption{\label{ususfesr}Mass prediction for the $us\bar{u}\bar{s}$ tetraquark state versus $s_0$ in FESR by using $J_1$ considering in the OPE series the perturbative term (yellow continuous), d=2 term (green dashed), d=4 term (red dotted), d=6 term (blue dotted-dashed), d=8 term (black continuous).
}
\label{fig:ususfesr}
\end{figure}

The situation for the $us\bar{u}\bar{s}$ tetraquark state is somewhat different from that for the $ud\bar{u}\bar{d}$ system. The OPE degeneracy between $J_1$ and $J_5$ (also for $J_2$/$J_8$, $J_3$/$J_7$ or $J_4$/$J_6$) in the $ud\bar{u}\bar{d}$ system is slightly changed for the $us\bar{u}\bar{s}$ system due to the SU(3) breaking. The coefficients of $\GG \ms \qq$ have opposite signs for $J_1$ and $J_5$, as shown in Appendix \ref{app:A}. Although the contributions of these terms are small, we still perform numerical analyses for all interpolating currents in the $us\bar{u}\bar{s}$ system.  In this case LSR doesn't work as well as FESR: the LSR moment ratios are dominated by the dimension-6 condensates rather than the perturbative terms, which suggests that the OPE truncations are invalid. We shall take the current $J_1$ as an example. From FIG.~\ref{fig:ususlsrj3}, it is shown that the curves have the $s_0$ and $\tau$ stability. However, the OPE series are not converging but dominated by dimension-6 condensate terms, as shown in Table~\ref{tab:ususj3convergence}. This situation holds for all the other $us\bar{u}\bar{s}$ currents in LSR.
\begin{table}[htbp]
\renewcommand\arraystretch{1.5}
\centering
\caption{\label{tab:ususj3convergence}OPE terms at the LSR stability points using the $us\bar{u}\bar{s}$ currents.}
\begin{tabular}{ccccccccc}
  \hlinewd{.8pt}
    $J_i$ &${1\over\tau}{\hat B}\Pi_{i}^{d=0}/{\rm GeV}^4$ & ${1\over\tau}{\hat B}\Pi_{i}^{d=2}/{\rm GeV}^4$ & ${1\over\tau}{\hat B}\Pi_{i}^{d=4}/{\rm GeV}^4$ & ${1\over\tau}{\hat B}\Pi_{i}^{d=6}/{\rm GeV}^4$ & ${1\over\tau}{\hat B}\Pi_{i}^{d=8}/{\rm GeV}^4$ & ${1\over\tau}{\hat B}\Pi_{i}^{d=8}$/OPE & $\tau$/${\rm GeV}^{-2}$ & $s_0/{\rm GeV}^2$ \\
    \hline
  $J_1$ &$1.13112\times 10^{-6}$ &$-2.2152\times 10^{-7}$ &$-1.14284\times 10^{-6}$ &$-5.26536\times 10^{-6} $&$2.23076\times 10^{-7}$ &-0.0438063 &0.651 &6.17\\
  $J_2$ &$2.80721\times 10^{-6}$ &$-4.84744\times 10^{-6} $&$-1.07504\times 10^{-6} $&$-8.99266\times 10^{-6} $&$-8.23584\times 10^{-8}$ &0.0105215 &0.574 &5.7\\
   $J_3$ & $1.0706\times 10^{-6}$& $-1.78751\times 10^{-7}$& $-6.0989\times 10^{-8}$&$-3.10259\times 10^{-6}$ & $-6.50102\times 10^{-8}$&$0.0278209$ &$0.555$&5.28 \\
  $J_4$ &$2.26533\times 10^{-5}$ &$-2.76002\times 10^{-6}$&$3.85241\times 10^{-6} $&$-2.56595\times 10^{-5} $&$-2.62288\times 10^{-6}$ &0.578144 &0.405 &4.76\\
  $J_5$ &$9.84444\times 10^{-7}$ &$-1.99608\times 10^{-7}$ &$-1.06617 \times 10^{-6} $&$-5.08568\times 10^{-6} $&$2.91861\times 10^{-7}$ &-0.0575079 &0.674 &6.18\\
  $J_6$ &$1.68658\times 10^{-5}$ &$-2.21217\times 10^{-6} $&$3.32406\times 10^{-6}$ &$-2.38351\times 10^{-5}$ &$-2.00034\times 10^{-6}$ &0.254568 &0.436 &4.62\\
  $J_7$ &$1.54953\times 10^{-6} $&$-2.35871\times 10^{-7} $&$-7.3373\times 10^{-8}$ &$-3.40304\times 10^{-6}$ &$-1.73889\times 10^{-7} $&0.0744181 &0.506 &5.28\\
  $J_8$ &$2.27458\times 10^{-6}$ &$-4.13982\times 10^{-7} $&$-9.67697\times 10^{-7} $&$-8.53188\times 10^{-6}$ &$8.61095\times 10^{-8}$ &-0.0114009 &0.605 &5.68\\
  \hlinewd{.8pt}
\end{tabular}
\end{table}
We then perform the FESR analyses for the $us\bar{u}\bar{s}$ systems. In our analysis, the FESR ratio shows nice behavior in the sense that it reach stability in the continuum threshold $s_0$, as shown in Fig.~\ref{fig:ususfesr}. The dimension-8 condensate only slightly affect the mass prediction, which justify the validity of OPE truncation. The same situations happen for all the other currents $J_2-J_8$. The mass predictions obtained from the stability points read
%%%%%%%%%%%%%%%%%%%%%%%%%%%%%%%1
%%%%%%%%%%%%%%%%%%%%%%%%%%%%%%%%%%%%
\begin{eqnarray}
M_{J_{1};FESR}&=&1.65(1.69)~{\rm GeV}~at~s_0=6.17(6.45)~{\rm GeV^2},\nnb\\
M_{J_{2};FESR}&=&1.58(1.63)~{\rm GeV}~at~s_0=5.67(5.99)~{\rm GeV^2},\nnb\\
M_{J_{3};FESR}&=&1.51(1.57)~{\rm GeV}~at~s_0=5.25(5.60)~{\rm GeV^2},\nnb\\
M_{J_{4};FESR}&=&1.44(1.51)~{\rm GeV}~at~s_0=4.76(5.15)~{\rm GeV^2},\nnb\\  \label{ususmass}
M_{J_{5};FESR}&=&1.65(1.69)~{\rm GeV}~at~s_0=6.16(6.44)~{\rm GeV^2},\\ 
M_{J_{6};FESR}&=&1.43(1.51)~{\rm GeV}~at~s_0=4.75(5.14)~{\rm GeV^2},\nnb\\
M_{J_{7};FESR}&=&1.51(1.57)~{\rm GeV}~at~s_0=5.26(5.61)~{\rm GeV^2},\nnb\\
\nonumber
M_{J_{8};FESR}&=&1.57(1.62)~{\rm GeV}~at~s_0=5.66(5.98)~{\rm GeV^2}.
\end{eqnarray}
We finally give the mass of the hidden-strange $0^{+-}$ tetraquark to be
\begin{eqnarray}
M_{us\bar{u}\bar{s}}&=&1.54\pm0.12~{\rm GeV}\, ,
\end{eqnarray}
where we have already considered the mass differences in Eq.~\eqref{ususmass} as one kind of the error sources.
%%%%%%%%%%%%%%%%%%%%%%%%%%%%%%%%%%%%%%%%
\section{Decay patterns of the $0^{+-}$ tetraquarks}
%%%%%%%%%%%%%%%%%%%%%%%%%%%%%%%%%%%%%%%%
Under the SU(2) symmetry, the $ud\bar u\bar d$ interpolating currents would couple to all $I=0, 1, 2$ isospin multiplets. In our calculation, these multiplets shall degenerate since we do not consider the effects of isospin symmetry breaking. In other words, our calculations will give the same mass predictions for the $I^GJ^{PC}=0^-0^{+-}, 1^+0^{+-}, 2^-0^{+-}$ tetraquarks with quark contents $ud\bar u\bar d$. For the $us\bar{u}\bar{s}$ tetraquarks, the quantum numbers will be $I^GJ^{PC}=0^-0^{+-}, 1^+0^{+-}$ considering the isospin. Tetraquark states have abundant decay modes so long as the kinematics allows. 

Considering the symmetry constraints for the isospin $I$, spin $J$, parity $P$, $C$-parity and $G$-parity, all S-wave two-body strong decays are forbidden for the charge neutral $ud\bar u \bar d$ tetraquarks, while only P-wave decay modes are allowed (as shown in Table~\ref{ududdecay}). Their dominant decay modes $b_1\pi, h_1\pi, a_1\pi$ are suitable for detection and partial-wave analyses. The charged partners $ud\bar d \bar d$ and $ud\bar u \bar u$ in the isospin-1 multiplet can dominantly decay into $\pi\pi$ modes in $S$-wave, and thus are expected to be very wide. 
%%%%%%%%%%%%%%%%%%%%%
\begin{center}
\renewcommand{\arraystretch}{1.3}
\begin{tabular*}{5.6cm}{ccc}
%\begin{tabular}{ccc}
\hlinewd{.8pt}
~~~$I^GJ^{PC}$ ~~~ & ~~~     $S$-wave            ~~~         &        ~~~      $P$-wave                \\
\hline
$0^-0^{+-}$  & $\times$                                                                     & $b_1\pi\, , \sigma\omega$ \\
$1^+0^{+-}$  & $\times$                                                    & $h_1\pi\, , \rho\sigma\, , a_1\pi$\\
$2^-0^{+-}$  & $\times$                                                    & $b_1\pi$\\
\hlinewd{.8pt}
%\end{tabular}
\end{tabular*}
\tabcaption{Possible two-body strong decay modes for the charge neutral $ud\bar u \bar d$ tetraquarks. \label{ududdecay}}
\end{center}
%%%%%%%%%%%%%%%%%%%%%%%%%%%%%%%%%%%%%%%%%%%%%%%%%%%%%%%%%

For the isovector $u s\bar u \bar s$ tetraquark, both two-body and three-body strong decays are totally forbidden by the kinematics and the 
strong constraints on the exotic quantum numbers $I^GJ^{PC}=1^+0^{+-}$. Especially the $S$-wave $K^+K^-$ and $K^0\bar K^0$ final states are forbidden by the negative $C$-parity of the tetraquarks. In other words, the charge neutral isovector $us\bar u \bar s$ tetraquark state can only decay via weakly interaction into final states such as $K\pi\pi$. It is predicted to be very narrow. 
For the charged partners $us\bar d \bar s$ and $ds\bar u \bar s$ states, one may also search for them in the $K^+\bar K^0, K^-K^0$ hadronic modes respectively. 

The decay behavior of the isoscalar $u s\bar u \bar s$ tetraquark will be different. Besides the weak decays, the dominant decay mode for the isoscalar $us\bar u \bar s$ tetraquark will be the three-body hadronic mode $us\bar u \bar s\to\phi\pi\pi  (\sigma\to\pi\pi)$. However, such a three-body decay is relatively small due to a $P$-wave suppression between $\phi$ and $\sigma (\pi\pi)$. This isoscalar $u s\bar u \bar s$ tetraquark state may be not very wide $\Gamma<1$ MeV and has convenient decay modes like $\phi\pi\pi, K\pi\pi$.
%%%%%%%%%%%%%%%%%%%%%%%%%%%%%%%%%%%%%%%%
\section{Summary and conclusions}
%%%%%%%%%%%%%%%%%%%%%%%%%%%%%%%%%%%%%%%%

In this work, we have studied the $ud\bar{u}\bar{d}$ and $us\bar{u}\bar{s}$ tetraquark states with $J^{PC}=0^{+-}$ in Laplace sum rules and finite energy sum rules. We consider all possible diquark-antidiquark interpolating currents with such exotic quantum numbers without containing covariant derivative operators. The dimension of such currents is lower than those used in Ref.~\cite{Du:2012pn}, which will result in better OPE behaviors and more reliable mass predictions.

We find that both LSR and FESR work well for the $ud\bar{u}\bar{d}$ system following the standard stability criteria of QCD sum rules and the requirement of well justified OPE truncation. For the $us\bar{u}\bar{s}$ system, all interpolating currents suffer from poor OPE convergence in LSR. However, the FESR ratios could yield to good $s_0$-stability, around which one can extract the hadronic masses. Considering both the LSR and FESR results, we predict the masses of the $ud\bar{u}\bar{d}$ and the $us\bar{u}\bar{s}$ tetraquarks to be $1.43\pm0.09~{\rm GeV}$ and $1.54\pm0.12~{\rm GeV},$ respectively.

Accordingly, we discuss the possible decay patterns for these exotic tetraquarks. The light $ud\bar{u}\bar{d}$ tetraquark can decay into two-body final states in P-wave hadronic modes, as shown in Table~\ref{ududdecay}.
Our analyses show that the $0^{+-}$ charge neutral isovector $us\bar{u}\bar{s}$ tetraquark may only decay via the weak interaction mechanism such as $us\bar{u}\bar{s}\to K\pi\pi$, since its strong decay modes are all forbidden by the kinematics and strong constraints on the exotic quantum numbers. It is predicted to be very narrow. The $0^{+-}$ isoscalar  $us\bar{u}\bar{s}$ tetraquark is also expected to be not very wide due to its $P$-wave dominant decay mode $us\bar u \bar s\to\phi\pi\pi$. These tetraquark states may be detectable in the near future at BESIII and Belle~II.

%%%%%%%%%%%%%%%%%%%%%%%%%%%%%%
\section*{ACKNOWLEDGMENTS}
This project is supported by the Chinese National Youth Thousand
Talents Program and the China Postdoctoral Science Foundation funded project (2018M631572).

\begin{appendix}
\clearpage
\section{The coefficients of LSR moments for the $ud\bar{u}\bar{d}$ and $us\bar{u}\bar{s}$ currents }\label{app:A}
In this appendix, we present the Wilson coefficients in Eqs.~\eqref{eq:udud} (see Eqs.~\eqref{eq:ududope1}-\eqref{eq:ududope8}) and \eqref{eq:usus} (see Eqs.~\eqref{eq:ope1}-\eqref{eq:ope8}).%(see TABLE~\ref{tab:coeff})
%\begin{table}[htbp]
%\centering
%\caption{\label{tab:coeff} Coefficients in Eq.~\eqref{eq:udud}.}
%\begin{tabular}{|c|c|c|c|c|c|c|c|c|}
%\hline
%\multirow{2}{*}{}& \multicolumn{8}{c|}{$i$}
%  \\
%  \cline{2-9}
%   & 1 & 2 & 3 & 4 & 5 & 6 & 7 & 8 \\
%  \hline

% $a_{i}$ & 1/30720 &  1/20480 &1/61440 & 1/10240 & 1/30720 & 1/10240 & 1/61440 & 1/20480 \\
%  \hline
%  $b_{i}$ & -1/1536 & 1/1536 &1/1536 & 11/1536 & -1/1536 & 11/1536 & 1/1536 & 1/1536 \\
%  \hline
%  $c_{i}$ & 1/6  &1/4  &1/12  & 1/2 & 1/6 & 1/2 & 1/12 & 1/4 \\
 % \hline
%  $d_{i}$ &1/24 &1/16  & 1/48 & 1/8 & 1/24 & 1/8 & 1/48 & 1/16 \\
%  \hline
%\end{tabular}
%\end{table}
\begin{flalign}
\label{eq:ududope1}
a_1&=\frac{1}{30270\pi^6}, b_1=\frac{-\langle \alpha_{s}G^2\rangle}{1536\pi^5}, c_1=\frac{\langle \bar qq\rangle^2}{6\pi^2}, d_1=\frac{\langle \bar qGq\rangle \langle \bar qq\rangle}{12\pi^2}\\
a_2&=\frac{1}{20480\pi^6}, b_2=\frac{\langle \alpha_{s}G^2\rangle}{1536\pi^5}, c_2=\frac{\langle \bar qq\rangle^2}{4\pi^2}, d_2=\frac{\langle \bar qGq\rangle \langle \bar qq\rangle}{8\pi^2}\\
a_3&=\frac{1}{61440\pi^6}, b_3=\frac{\langle \alpha_{s}G^2\rangle}{1536\pi^5}, c_3=\frac{\langle \bar qq\rangle^2}{12\pi^2}, d_3=\frac{\langle \bar qGq\rangle \langle \bar qq\rangle}{24\pi^2}\\
a_4&=\frac{1}{10240\pi^6}, b_4=\frac{11\langle \alpha_{s}G^2\rangle}{1536\pi^5}, c_4=\frac{\langle \bar qq\rangle^2}{2\pi^2}, d_4=\frac{\langle \bar qGq\rangle \langle \bar qq\rangle}{4\pi^2}\\
a_5&=\frac{1}{30270\pi^6}, b_5=\frac{-\langle \alpha_{s}G^2\rangle}{1536\pi^5}, c_5=\frac{\langle \bar qq\rangle^2}{6\pi^2}, d_5=\frac{\langle \bar qGq\rangle \langle \bar qq\rangle}{12\pi^2}\\
a_6&=\frac{1}{10240\pi^6}, b_6=\frac{11\langle \alpha_{s}G^2\rangle}{1536\pi^5}, c_6=\frac{\langle \bar qq\rangle^2}{2\pi^2}, d_6=\frac{\langle \bar qGq\rangle \langle \bar qq\rangle}{4\pi^2}\\
a_7&=\frac{1}{61440\pi^6}, b_7=\frac{\langle \alpha_{s}G^2\rangle}{1536\pi^5}, c_7=\frac{\langle \bar qq\rangle^2}{12\pi^2}, d_7=\frac{\langle \bar qGq\rangle \langle \bar qq\rangle}{24\pi^2}\\
a_8&=\frac{1}{20480\pi^6}, b_8=\frac{\langle \alpha_{s}G^2\rangle}{1536\pi^5}, c_8=\frac{\langle \bar qq\rangle^2}{4\pi^2}, d_8=\frac{\langle \bar qGq\rangle \langle \bar qq\rangle}{8\pi^2}
\label{eq:ududope8}
\end{flalign}

\begin{flalign}
a_1^{\prime}&=-\frac{1}{30720 \pi ^6},
b_1^{\prime}=\frac{5 \ms^2}{1536 \pi ^6},
c_1^{\prime}=\frac{\GG}{1536 \pi ^5}-\frac{\ms^4}{256 \pi ^6}-\frac{\ms \ss}{32 \pi ^4},
d_1^{\prime}=-\frac{\GG \ms^2}{384 \pi ^5}-\frac{\ms \sGs}{32 \pi ^4}+\frac{\qq ^2}{12 \pi ^2}+\frac{\ss^2}{12 \pi ^2},\nnb \hspace{36pt}\\
e_1^{\prime}&=\lbrack 2+\gamma_{E}-\ln(4 \pi) \rbrack \ga \frac{\ms^2 \qq^2}{6 \pi ^2}+\frac{\ms^2 \ss^2}{24 \pi ^2}+\frac{\GG \ms \qq}{576 \pi ^3}-\frac{\GG \ms \ss}{144 \pi ^3}-\frac{\qGq \qq}{24 \pi ^2}-\frac{\sGs \ss}{24 \pi ^2} \dr, \hspace{36pt}\nnb\\
f_1^{\prime}&=-\frac{\ms^2 \qq^2}{2 \pi ^2}+\frac{\ms^2 \ss^2}{24 \pi ^2}-\frac{\GG \ms \qq}{64 \pi ^3}+\frac{\qGq \qq}{24 \pi ^2}+\frac{\sGs \ss}{24 \pi ^2} \hspace{36pt}
  \label{eq:ope1}
\end{flalign}

 %-\frac{\gamma_E  \ms^2 \qq ^2}{6 \pi ^2}+\frac{\ms^2 \qq^2 Log(2) }{3 \pi ^2}+\frac{\ms^2 \qq^2 Log(\pi )}{6 \pi ^2}-\frac{\ms^2 \ss^2}{12 \pi ^2}-\frac{\gamma_E \ms^2 \ss ^2}{24 \pi ^2}+\frac{\ms^2 \ss^2 Log(2)}{12 \pi ^2}+\frac{\ms^2 \ss^2 Log(\pi)}{24 \pi ^2}\nonumber\\
  %&-\frac{\GG \ms \qq}{288 \pi ^3}-\frac{\gamma_E  \GG \ms \qq}{576 \pi ^3}+\frac{\GG \ms \qq Log(2)}{288 \pi ^3}+\frac{\GG \ms \qq Log(\pi)}{576 \pi ^3}+\frac{\GG \ms \ss}{144 \pi ^3}+\frac{\gamma_E  \GG \ms \ss}{288 \pi ^3}\nonumber\\
  %&-\frac{\GG \ms \ss Log(2)}{144 \pi ^3}-\frac{\GG \ms \ss Log(\pi)}{288 \pi ^3}+\frac{\qGq \qq}{12 \pi ^2}+\frac{\gamma_E  \qGq \qq}{24 \pi ^2}-\frac{\qGq \qq Log(2)}{12 \pi ^2}-\frac{\qGq \qq Log(\pi)}{24 \pi ^2}\nonumber\\
  %&+\frac{\sGs \ss}{12 \pi ^2}+\frac{\gamma_E  \sGs \ss}{24 \pi ^2}-\frac{\sGs \ss Log(2)}{12 \pi ^2}-\frac{\sGs\ss Log(\pi)}{24 \pi ^2}

\begin{flalign}
a_2^{\prime}&=-\frac{1}{20480 \pi ^6},
b_2^{\prime}=\frac{5 \ms^2}{1024 \pi ^6},
c_2^{\prime}=-\frac{\GG}{1536 \pi ^5}-\frac{3 \ms^4}{512 \pi ^6}-\frac{3 \ms \ss}{64 \pi ^4},
d_2^{\prime}=\frac{3 \GG \ms^2}{512 \pi ^5}-\frac{3 \ms \sGs}{64 \pi ^4}+\frac{\qq^2}{8 \pi ^2}+\frac{\ss^2}{8 \pi ^2},\nnb \\
e_2^{\prime}&=\lbrack 2+\gamma_{E}-\ln(4 \pi) \rbrack \ga \frac{\GG \ms^4}{284 \pi ^5}+\frac{\ms^2 \qq^2}{4 \pi ^2}+\frac{\ms^2 \ss^2}{16 \pi ^2}+\frac{\GG \ms \qq}{288 \pi ^3}+\frac{\GG \ms \ss}{192 \pi ^3}-\frac{\qGq \qq}{16 \pi ^2}-\frac{\sGs \ss}{16 \pi ^2} \dr,\nnb \\
f_2^{\prime}&=-\frac{3 \ms^2 \qq^2}{4 \pi ^2}+\frac{\ms^2 \ss^2}{16 \pi ^2}-\frac{\GG \ms \qq}{32 \pi ^3}-\frac{\GG \ms \ss}{64 \pi ^3}+\frac{\qGq \qq}{16 \pi ^2}+\frac{\sGs \ss}{16 \pi ^2}
\end{flalign}

%-\frac{\GG \ms^4}{192 \pi ^5}-\frac{\gamma_E  \GG \ms^4}{384 \pi ^5}+\frac{\GG \ms^4 Log(2)}{192 \pi ^5}+\frac{\GG \ms^4 Log(\pi)}{384 \pi ^5}-\frac{\ms^2 \qq^2}{2 \pi ^2}-\frac{\gamma_E  \ms^2 \qq^2}{4 \pi ^2}+\frac{\ms^2 \qq^2 Log(2)}{2 \pi ^2}\nonumber\\
%&+\frac{\ms^2 \qq^2 Log(\pi)}{4 \pi ^2}-\frac{\ms^2 \ss^2}{8 \pi ^2}-\frac{\gamma_E  \ms^2 \ss^2}{16 \pi ^2}+\frac{\ms^2 \ss^2 Log(2)}{8 \pi ^2}+\frac{\ms^2 \ss^2 Log(\pi)}{16 \pi ^2}-\frac{\GG \ms \qq}{144 \pi ^3}-\frac{\gamma_E  \GG \ms \qq}{288 \pi ^3}\nonumber\\
%&+\frac{\GG \ms \qq Log(2)}{144 \pi ^3}+\frac{\GG \ms \qq Log(\pi)}{288 \pi ^3}-\frac{\GG \ms \ss}{96 \pi ^3}-\frac{\gamma_E  \GG \ms \ss}{192 \pi ^3}+\frac{\GG \text{ms} \ss Log(2)}{96 \pi ^3}\nonumber\\
%&+\frac{\GG \ms \ss Log(\pi)}{192 \pi ^3}+\frac{\qGq \qq}{8 \pi ^2}+\frac{\gamma_E  \qGq \qq}{16 \pi ^2}-\frac{\qGq \qq Log(2)}{8 \pi ^2}-\frac{\qGq \qq Log(\pi)}{16 \pi ^2}+\frac{\sGs \ss}{8 \pi ^2}+\frac{\gamma_E  \sGs \ss}{16 \pi ^2}\nonumber\\
%&-\frac{\sGs \ss Log(2)}{8 \pi ^2}-\frac{\sGs \ss Log(\pi)}{16 \pi ^2}

 \begin{flalign}
a_3^{\prime}&=-\frac{1}{61440 \pi ^6},
b_3^{\prime}=\frac{5 \ms^2}{3072 \pi ^6},
c_3^{\prime}=-\frac{\GG}{1536 \pi ^5}-\frac{\ms^4}{512 \pi ^6}-\frac{\ms \ss}{64 \pi ^4},
d_3^{\prime}=\frac{\GG \ms^2}{384 \pi ^5}-\frac{\ms \sGs}{64 \pi ^4}+\frac{\qq^2}{24 \pi ^2}+\frac{\ss^2}{24 \pi ^2},\nonumber \hspace{35pt}\\
e_3^{\prime}&=\lbrack 2+\gamma_{E}-\ln(4 \pi) \rbrack \ga \frac{\ms^2 \qq^2}{12 \pi ^2}+\frac{\ms^2 \ss^2}{48 \pi ^2}-\frac{\GG \ms \qq}{576 \pi ^3}+\frac{\GG \ms \ss}{288 \pi ^3}-\frac{\qGq \qq}{48 \pi ^2}-\frac{\sGs \ss}{48 \pi ^2} \dr, \hspace{35pt} \nnb\\
f_3^{\prime}&=-\frac{\ms^2 \qq^2}{4 \pi ^2}+\frac{\ms^2 \ss^2}{48 \pi ^2}+\frac{\GG \ms \qq}{64 \pi ^3}+\frac{\qGq \qq}{48 \pi ^2}+\frac{\sGs \ss}{48 \pi ^2} \hspace{35pt}
 \end{flalign}
%-\frac{\ms^2 \qq^2}{6 \pi ^2}-\frac{\gamma_E  \ms^2 \qq^2}{12 \pi ^2}+\frac{\ms^2 \qq^2 Log(2)}{6 \pi ^2}+\frac{\ms^2 \qq^2 Log(\pi)}{12 \pi ^2}-\frac{\ms^2 \ss^2}{24 \pi ^2}-\frac{\gamma_E  \ms^2 \ss^2}{48 \pi ^2}+\frac{\ms^2 \ss^2 Log(2)}{24 \pi ^2}+\frac{\ms^2 \ss^2 Log(\pi)}{48 \pi ^2}\nonumber\\
%&+\frac{\GG \ms \qq}{288 \pi ^3}+\frac{\gamma_E  \GG \ms \qq}{576 \pi ^3}-\frac{\GG \ms \qq Log(2)}{288 \pi ^3}-\frac{\GG \ms \qq Log(\pi)}{576 \pi ^3}-\frac{\GG \ms \ss}{144 \pi ^3}-\frac{\gamma_E  \GG \ms \ss}{288 \pi ^3}\nonumber\\
%&+\frac{\GG \ms \ss Log(2)}{144 \pi ^3}+\frac{\GG \ms \ss Log(\pi)}{288 \pi ^3}+\frac{\qGq \qq}{24 \pi ^2}+\frac{\gamma_E  \qGq \qq}{48 \pi ^2}-\frac{\qGq \qq Log(2)}{24 \pi ^2}-\frac{\qGq \qq Log(\pi)}{48 \pi ^2}\nonumber\\
%&+\frac{\sGs \ss}{24 \pi ^2}+\frac{\gamma_E  \sGs \ss}{48 \pi ^2}-\frac{\sGs \ss Log(2)}{24 \pi ^2}-\frac{\sGs \ss Log(\pi)}{48 \pi ^2}

 \begin{flalign}
a_4^{\prime}&=-\frac{1}{10240 \pi ^6},
b_4^{\prime}=\frac{5 \ms^2}{512 \pi ^6},
c_4^{\prime}=-\frac{11 \GG}{1536 \pi ^5}-\frac{3 \ms^4}{256 \pi ^6}-\frac{3 \ms \ss}{32 \pi ^4},
d_4^{\prime}=\frac{23 \GG \ms^2}{512 \pi ^5}-\frac{3 \ms \sGs}{32 \pi ^4}+\frac{\qq^2}{4 \pi ^2}+\frac{\ss^2}{4 \pi ^2},\nonumber\\
e_4^{\prime}&=\lbrack 2+\gamma_{E}-\ln(4 \pi) \rbrack \ga \frac{5\GG \ms^4}{384 \pi ^5}+\frac{\ms^2 \qq^2}{2 \pi ^2}+\frac{\ms^2 \ss^2}{8 \pi ^2}+\frac{\GG \ms \qq}{144 \pi ^3}+\frac{\GG \ms \ss}{64 \pi ^3}-\frac{\qGq \qq}{8 \pi ^2}-\frac{\sGs \ss}{8 \pi ^2} \dr,\nnb \\
f_4^{\prime}&=-\frac{3 \ms^2 \qq^2}{2 \pi ^2}+\frac{\ms^2 \ss^2}{8 \pi ^2}-\frac{\GG \ms \qq}{16 \pi ^3}-\frac{5 \GG \ms \ss}{64 \pi ^3}+\frac{\qGq \qq}{8 \pi ^2}+\frac{\sGs \ss}{8 \pi ^2}
 \end{flalign}
%-\frac{5 \GG \ms^4}{192 \pi ^5}-\frac{5 \gamma_E  \GG \ms^4}{384 \pi ^5}+\frac{5 \GG \ms^4 Log(2)}{192 \pi ^5}+\frac{5 \GG \ms^4 Log(\pi)}{384 \pi ^5}-\frac{\ms^2 \qq^2}{\pi ^2}-\frac{\gamma_E \ms^2 \qq^2}{2 \pi ^2}+\frac{\ms^2 \qq^2 Log(2)}{\pi ^2}\nonumber\\
%&+\frac{\ms^2 \qq^2 Log(\pi)}{2 \pi ^2}-\frac{\ms^2 \ss^2}{4 \pi ^2}-\frac{\gamma_E  \ms^2 \ss^2}{8 \pi ^2}+\frac{\ms^2 \ss^2 Log(2)}{4 \pi ^2}+\frac{\ms^2 \ss^2 Log(\pi)}{8 \pi ^2}-\frac{\GG \ms \qq}{72 \pi ^3}-\frac{\gamma_E  \GG \ms \qq}{144 \pi ^3}\nonumber\\
%&+\frac{\GG \ms \qq Log(2)}{72 \pi ^3}+\frac{\GG \ms \qq Log(\pi)}{144 \pi ^3}-\frac{3 \GG \ms \ss}{32 \pi ^3}-\frac{3 \gamma_E  \GG \ms \ss}{64 \pi ^3}+\frac{3 \GG \ms\ss Log(2)}{32 \pi ^3}\nonumber\\
%&+\frac{3 \GG \ms \ss Log(\pi)}{64 \pi ^3}+\frac{\qGq \qq}{4 \pi ^2}+\frac{\gamma_E  \qGq \qq}{8 \pi ^2}-\frac{\qGq \qq Log(2)}{4 \pi ^2}-\frac{\qGq \qq Log(\pi)}{8 \pi ^2}+\frac{\sGs \ss}{4 \pi ^2}+\frac{\gamma_E  \sGs \ss}{8 \pi ^2}\nonumber\\
%&-\frac{\sGs \ss Log(2)}{4 \pi ^2}-\frac{\sGs \ss Log(\pi)}{8 \pi ^2}

 \begin{flalign}
a_5^{\prime}&=-\frac{1}{30720 \pi ^6},
b_5^{\prime}=\frac{5 \ms^2}{1536 \pi ^6},
c_5^{\prime}=\frac{\GG}{1536 \pi ^5}-\frac{\ms^4}{256 \pi ^6}-\frac{\ms \ss}{32 \pi ^4},
d_5^{\prime}=-\frac{\GG \ms^2}{384 \pi ^5}-\frac{\ms \sGs}{32 \pi ^4}+\frac{\qq^2}{12 \pi ^2}+\frac{\ss^2}{12 \pi ^2},\nonumber \hspace{41pt}\\
e_5^{\prime}&=\lbrack 2+\gamma_{E}-\ln(4 \pi) \rbrack \ga \frac{\ms^2 \qq^2}{6 \pi ^2}+\frac{\ms^2 \ss^2}{24 \pi ^2}-\frac{\GG \ms \qq}{576 \pi ^3}-\frac{\GG \ms \ss}{288 \pi ^3}-\frac{\qGq \qq}{24 \pi ^2}-\frac{\sGs \ss}{24 \pi ^2} \dr, \hspace{41pt}\nnb\\
f_5^{\prime}&=-\frac{\ms^2 \qq^2}{2 \pi ^2}+\frac{\ms^2 \ss^2}{24 \pi ^2}+\frac{\GG \ms \qq}{64 \pi ^3}+\frac{\qGq \qq}{24 \pi ^2}+\frac{\sGs \ss}{24 \pi ^2} \hspace{41pt}
 \end{flalign}
%-\frac{\ms^2 \qq^2}{3 \pi ^2}-\frac{\gamma_E  \ms^2 \qq^2}{6 \pi ^2}+\frac{\ms^2 \qq^2 Log(2)}{3 \pi ^2}+\frac{\ms^2 \qq^2 Log(\pi)}{6 \pi ^2}-\frac{\ms^2 \ss^2}{12 \pi ^2}-\frac{\gamma_E  \ms^2 \ss^2}{24 \pi ^2}+\frac{\ms^2 \ss^2 Log(2)}{12 \pi ^2}+\frac{\ms^2 \ss^2 Log(\pi)}{24 \pi ^2}\nonumber\\
%&+\frac{\GG \ms \qq}{288 \pi ^3}+\frac{\gamma_E  \GG \ms \qq}{576 \pi ^3}-\frac{\GG \ms \qq Log(2)}{288 \pi ^3}-\frac{\GG \ms \qq Log(\pi)}{576 \pi ^3}+\frac{\GG \ms \ss}{144 \pi ^3}+\frac{\gamma_E  \GG \ms \ss}{288 \pi ^3}\nonumber\\
%&-\frac{\GG \ms \ss Log(2)}{144 \pi ^3}-\frac{\GG \ms \ss Log(\pi)}{288 \pi ^3}+\frac{\qGq \qq}{12 \pi ^2}+\frac{\gamma_E  \qGq \qq}{24 \pi ^2}-\frac{\qGq \qq Log(2)}{12 \pi ^2}-\frac{\qGq \qq Log(\pi)}{24 \pi ^2}\nonumber\\
%&+\frac{\sGs \ss}{12 \pi ^2}+\frac{\gamma_E  \sGs \ss}{24 \pi ^2}-\frac{\sGs \ss Log(2)}{12 \pi ^2}-\frac{\sGs \ss Log(\pi)}{24 \pi ^2}

 \begin{flalign}
a_6^{\prime}&=-\frac{1}{10240 \pi ^6},
b_6^{\prime}=\frac{5 \ms^2}{512 \pi ^6},
c_6^{\prime}=-\frac{11 \GG}{1536 \pi ^5}-\frac{3 \ms^4}{256 \pi ^6}-\frac{3 \ms \ss}{32 \pi ^4},
d_6^{\prime}=\frac{23 \GG \ms^2}{512 \pi ^5}-\frac{3 \ms \sGs}{32 \pi ^4}+\frac{\qq^2}{4 \pi ^2}+\frac{\ss^2}{4 \pi ^2},\nonumber\\
e_6^{\prime}&=\lbrack 2+\gamma_{E}-\ln(4 \pi) \rbrack \ga \frac{5\GG \ms^4}{384 \pi ^5}+\frac{\ms^2 \qq^2}{2 \pi ^2}+\frac{\ms^2 \ss^2}{8 \pi ^2}-\frac{\GG \ms \qq}{144 \pi ^3}+\frac{\GG \ms \ss}{64 \pi ^3}-\frac{\qGq \qq}{8 \pi ^2}-\frac{\sGs \ss}{8 \pi ^2} \dr,\nnb\\
f_6^{\prime}&=-\frac{3 \ms^2 \qq^2}{2 \pi ^2}+\frac{\ms^2 \ss^2}{8 \pi ^2}+\frac{\GG \ms \qq}{16 \pi ^3}-\frac{5 \GG \ms \ss}{64 \pi ^3}+\frac{\qGq \qq}{8 \pi ^2}+\frac{\sGs \ss}{8 \pi ^2}
\end{flalign}
%-\frac{5 \GG \ms^4}{192 \pi ^5}-\frac{5 \gamma_E  \GG \ms^4}{384 \pi ^5}+\frac{5 \GG \ms^4 Log(2)}{192 \pi ^5}+\frac{5 \GG \ms^4 Log(\pi)}{384 \pi ^5}-\frac{\ms^2 \qq^2}{\pi ^2}-\frac{\gamma_E  \ms^2 \qq^2}{2 \pi ^2}+\frac{\ms^2 \qq^2 Log(2)}{\pi ^2}\nonumber\\
%&+\frac{\ms^2 \qq^2 Log(\pi)}{2 \pi ^2}-\frac{\ms^2 \ss^2}{4 \pi ^2}-\frac{\gamma_E  \ms^2 \ss^2}{8 \pi ^2}+\frac{\ms^2 \ss^2 Log(2)}{4 \pi ^2}+\frac{\ms^2 \ss^2 Log(\pi)}{8 \pi ^2}+\frac{\GG \ms \qq}{72 \pi ^3}+\frac{\gamma_E  \GG \ms \qq}{144 \pi ^3}\nonumber\\
%&-\frac{\GG \ms \qq Log(2)}{72 \pi ^3}-\frac{\GG \ms \qq Log(\pi)}{144 \pi ^3}-\frac{3 \GG \ms \ss}{32 \pi ^3}-\frac{3 \gamma_E  \GG \ms \ss}{64 \pi ^3}+\frac{3 \GG \ms \ss Log(2)}{32 \pi ^3}\nonumber\\
%&+\frac{3 \GG \ms \ss Log(\pi)}{64 \pi ^3}+\frac{\qGq \qq}{4 \pi ^2}+\frac{\gamma_E  \qGq \qq}{8 \pi ^2}-\frac{\qGq \qq Log(2)}{4 \pi ^2}-\frac{\qGq \qq Log(\pi)}{8 \pi ^2}+\frac{\sGs \ss}{4 \pi ^2}+\frac{\gamma_E  \sGs \ss}{8 \pi ^2}\nonumber\\
%&-\frac{\sGs \ss Log(2)}{4 \pi ^2}-\frac{\sGs \ss Log(\pi)}{8 \pi ^2}

 \begin{flalign}
a_7^{\prime}&=-\frac{1}{61440 \pi ^6},
b_7^{\prime}=\frac{5 \ms^2}{3072 \pi ^6},
c_7^{\prime}=-\frac{\GG}{1536 \pi ^5}-\frac{\ms^4}{512 \pi ^6}-\frac{\ms \ss}{64 \pi ^4},
d_7^{\prime}=\frac{\GG \ms^2}{384 \pi ^5}-\frac{\ms \sGs}{64 \pi ^4}+\frac{\qq^2}{24 \pi ^2}+\frac{\ss^2}{24 \pi ^2},\nonumber \hspace{37pt}\\
e_7^{\prime}&=\lbrack 2+\gamma_{E}-\ln(4 \pi) \rbrack \ga \frac{\ms^2 \qq^2}{12 \pi ^2}+\frac{\ms^2 \ss^2}{48 \pi ^2}+\frac{\GG \ms \qq}{576 \pi ^3}+\frac{\GG \ms \ss}{288 \pi ^3}-\frac{\qGq \qq}{48 \pi ^2}-\frac{\sGs \ss}{48 \pi ^2} \dr, \hspace{37pt}\nnb\\
f_7^{\prime}&=-\frac{\ms^2 \qq^2}{4 \pi ^2}+\frac{\ms^2 \ss^2}{48 \pi ^2}-\frac{\GG \ms \qq}{64 \pi ^3}+\frac{\qGq \qq}{48 \pi ^2}+\frac{\sGs \ss}{48 \pi ^2}\hspace{37pt}
\end{flalign}
%-\frac{\ms^2 \qq^2}{6 \pi ^2}-\frac{\gamma_E  \ms^2 \qq^2}{12 \pi ^2}+\frac{\ms^2 \qq^2 Log(2) }{6 \pi ^2}+\frac{\ms^2 \qq^2 Log(\pi) }{12 \pi ^2}-\frac{\ms^2 \ss^2}{24 \pi ^2}-\frac{\gamma_E  \ms^2 \ss^2}{48 \pi ^2}+\frac{\ms^2 \ss^2 Log(2) }{24 \pi ^2}+\frac{\ms^2 \ss^2 Log(\pi) }{48 \pi ^2}\nonumber\\
%&-\frac{\GG \ms \qq}{288 \pi ^3}-\frac{\gamma_E  \GG \ms \qq}{576 \pi ^3}+\frac{\GG \ms \qq Log(2)}{288 \pi ^3}+\frac{\GG \ms \qq Log(\pi)}{576 \pi ^3}-\frac{\GG \ms \ss}{144 \pi ^3}-\frac{\gamma_E  \GG \ms \ss}{288 \pi ^3}\nonumber\\
%&+\frac{\GG \ms \ss Log(2)}{144 \pi ^3}+\frac{\GG \ms \ss Log(\pi) }{288 \pi ^3}+\frac{\qGq \qq}{24 \pi ^2}+\frac{\gamma_E  \qGq \qq}{48 \pi ^2}-\frac{\qGq \qq Log(2) }{24 \pi ^2}-\frac{\qGq \qq Log(\pi)}{48 \pi ^2}\nonumber\\
%&+\frac{\sGs \ss}{24 \pi ^2}+\frac{\gamma_E  \sGs \ss}{48 \pi ^2}-\frac{\sGs \ss Log(2) }{24 \pi ^2}-\frac{\sGs \ss Log(\pi) }{48 \pi ^2}

  \begin{flalign}
a_8^{\prime}&=-\frac{1}{20480 \pi ^6},
b_8^{\prime}=\frac{5 \ms^2}{1024 \pi ^6},
c_8^{\prime}=-\frac{\GG}{1536 \pi ^5}-\frac{3 \ms^4}{512 \pi ^6}-\frac{3 \ms \ss}{64 \pi ^4},
d_8^{\prime}=\frac{3 \GG \ms^2}{512 \pi ^5}-\frac{3 \ms \sGs}{64 \pi ^4}+\frac{\qq^2}{8 \pi ^2}+\frac{\ss^2}{8 \pi ^2},\nonumber\\
e_8^{\prime}&=\lbrack 2+\gamma_{E}-\ln(4 \pi) \rbrack \ga \frac{\GG \ms^4}{384 \pi ^5}+\frac{\ms^2 \qq^2}{4 \pi ^2}+\frac{\ms^2 \ss^2}{16 \pi ^2}-\frac{\GG \ms \qq}{288 \pi ^3}+\frac{\GG \ms \ss}{192 \pi ^3}-\frac{\qGq \qq}{16 \pi ^2}-\frac{\sGs \ss}{16 \pi ^2} \dr,\nnb \\
f_8^{\prime}&=-\frac{3 \ms^2 \qq^2}{4 \pi ^2}+\frac{\ms^2 \ss^2}{16 \pi ^2}+\frac{\GG \ms \qq}{32 \pi ^3}-\frac{\GG \ms \ss}{64 \pi ^3}+\frac{\qGq \qq}{16 \pi ^2}+\frac{\sGs \ss}{16 \pi ^2}
\label{eq:ope8}
 \end{flalign}

%-\frac{\GG \ms^4}{192 \pi ^5}-\frac{\gamma_E  \GG \ms^4}{384 \pi ^5}+\frac{\GG \ms^4 Log(2)}{192 \pi ^5}+\frac{\GG \ms^4 Log(\pi) }{384 \pi ^5}-\frac{\ms^2 \qq^2}{2 \pi ^2}-\frac{\gamma_E  \ms^2 \qq^2}{4 \pi ^2}+\frac{\ms^2 \qq^2 Log(2)}{2 \pi ^2}\nonumber\\
%&+\frac{\ms^2 \qq^2 Log(\pi)}{4 \pi ^2}-\frac{\ms^2 \ss^2}{8 \pi ^2}-\frac{\gamma_E  \ms^2 \ss^2}{16 \pi ^2}+\frac{\ms^2 \ss^2 Log(2)}{8 \pi ^2}+\frac{\ms^2 \ss^2 Log(\pi)}{16 \pi ^2}+\frac{\GG \ms \qq}{144 \pi ^3}+\frac{\gamma_E  \GG \ms \qq}{288 \pi ^3}\nonumber\\
%&-\frac{\GG \ms \qq Log(2)}{144 \pi ^3}-\frac{\GG \ms \qq Log(\pi)}{288 \pi ^3}-\frac{\GG \ms \ss}{96 \pi ^3}-\frac{\gamma_E  \GG \ms \ss}{192 \pi ^3}+\frac{\GG \ms \ss Log(2)}{96 \pi ^3}\nonumber\\
%&+\frac{\GG \ms \ss Log(\pi)}{192 \pi ^3}+\frac{\qGq \qq}{8 \pi ^2}+\frac{\gamma_E  \qGq \qq}{16 \pi ^2}-\frac{\qGq \qq Log(2)}{8 \pi ^2}-\frac{\qGq \qq Log(\pi)}{16 \pi ^2}+\frac{\sGs \ss}{8 \pi ^2}+\frac{\gamma_E  \sGs \ss}{16 \pi ^2}\nonumber\\
%&-\frac{\sGs \ss Log(2)}{8 \pi ^2}-\frac{\sGs \ss Log(\pi)}{16 \pi ^2}

  \end{appendix}

%\bibliography{ref}

\begin{thebibliography}{44}
\expandafter\ifx\csname natexlab\endcsname\relax\def\natexlab#1{#1}\fi
\expandafter\ifx\csname bibnamefont\endcsname\relax
  \def\bibnamefont#1{#1}\fi
\expandafter\ifx\csname bibfnamefont\endcsname\relax
  \def\bibfnamefont#1{#1}\fi
\expandafter\ifx\csname citenamefont\endcsname\relax
  \def\citenamefont#1{#1}\fi
\expandafter\ifx\csname url\endcsname\relax
  \def\url#1{\texttt{#1}}\fi
\expandafter\ifx\csname urlprefix\endcsname\relax\def\urlprefix{URL }\fi
\providecommand{\bibinfo}[2]{#2}
\providecommand{\eprint}[2][]{\url{#2}}

\bibitem[{\citenamefont{Klempt and Zaitsev}(2007)}]{2007-Klempt-p1-202}
\bibinfo{author}{\bibfnamefont{E.}~\bibnamefont{Klempt}} \bibnamefont{and}
  \bibinfo{author}{\bibfnamefont{A.}~\bibnamefont{Zaitsev}},
  \bibinfo{journal}{Phys. Rept.} \textbf{\bibinfo{volume}{454}},
  \bibinfo{pages}{1} (\bibinfo{year}{2007}), \eprint{arXiv:0708.4016[hep-ph]}.

\bibitem[{\citenamefont{Tanabashi et~al.}(2018)\citenamefont{Tanabashi,
  Hagiwara, Hikasa, Nakamura, Sumino, Takahashi, Tanaka, Agashe, Aielli, Amsler
  et~al.}}]{PhysRevD.98.030001}
\bibinfo{author}{\bibfnamefont{M.}~\bibnamefont{Tanabashi}},
  \bibinfo{author}{\bibfnamefont{K.}~\bibnamefont{Hagiwara}},
  \bibinfo{author}{\bibfnamefont{K.}~\bibnamefont{Hikasa}},
  \bibinfo{author}{\bibfnamefont{K.}~\bibnamefont{Nakamura}},
  \bibinfo{author}{\bibfnamefont{Y.}~\bibnamefont{Sumino}},
  \bibinfo{author}{\bibfnamefont{F.}~\bibnamefont{Takahashi}},
  \bibinfo{author}{\bibfnamefont{J.}~\bibnamefont{Tanaka}},
  \bibinfo{author}{\bibfnamefont{K.}~\bibnamefont{Agashe}},
  \bibinfo{author}{\bibfnamefont{G.}~\bibnamefont{Aielli}},
  \bibinfo{author}{\bibfnamefont{C.}~\bibnamefont{Amsler}},
  \bibnamefont{et~al.} (\bibinfo{collaboration}{Particle Data Group}),
  \bibinfo{journal}{Phys. Rev. D} \textbf{\bibinfo{volume}{98}},
  \bibinfo{pages}{030001} (\bibinfo{year}{2018}).

\bibitem[{\citenamefont{Ho et~al.}(2018)\citenamefont{Ho, Berg, Chen, and
  Steele}}]{Ho:2018cat}
\bibinfo{author}{\bibfnamefont{J.}~\bibnamefont{Ho}},
  \bibinfo{author}{\bibfnamefont{R.}~\bibnamefont{Berg}},
  \bibinfo{author}{\bibfnamefont{W.}~\bibnamefont{Chen}}, 
    \bibinfo{author}{\bibfnamefont{D.}~\bibnamefont{Harnett}}, \bibnamefont{and}
  \bibinfo{author}{\bibfnamefont{T.~G.} \bibnamefont{Steele}}
  (\bibinfo{year}{2018}), \eprint{arXiv:1806.02465[hep-ph]}.

\bibitem[{\citenamefont{Zhang et~al.}(2014)\citenamefont{Zhang, Jin, and
  Steele}}]{Zhang:2013rya}
\bibinfo{author}{\bibfnamefont{Z.-f.} \bibnamefont{Zhang}},
  \bibinfo{author}{\bibfnamefont{H.-y.} \bibnamefont{Jin}}, \bibnamefont{and}
  \bibinfo{author}{\bibfnamefont{T.~G.} \bibnamefont{Steele}},
  \bibinfo{journal}{Chin. Phys. Lett.} \textbf{\bibinfo{volume}{31}},
  \bibinfo{pages}{051201} (\bibinfo{year}{2014}),
  \eprint{arXiv:1312.5432[hep-ph]}.

\bibitem[{\citenamefont{Huang et~al.}(2015)\citenamefont{Huang, Jin, and
  Zhang}}]{Huang:2014hya}
\bibinfo{author}{\bibfnamefont{Z.-R.} \bibnamefont{Huang}},
  \bibinfo{author}{\bibfnamefont{H.-Y.} \bibnamefont{Jin}}, \bibnamefont{and}
  \bibinfo{author}{\bibfnamefont{Z.-F.} \bibnamefont{Zhang}},
  \bibinfo{journal}{JHEP} \textbf{\bibinfo{volume}{04}}, \bibinfo{pages}{004}
  (\bibinfo{year}{2015}), \eprint{arXiv:1411.2224[hep-ph]}.

\bibitem[{\citenamefont{Huang et~al.}(2016)\citenamefont{Huang, Jin, Steele,
  and Zhang}}]{Huang:2016upt}
\bibinfo{author}{\bibfnamefont{Z.-R.} \bibnamefont{Huang}},
  \bibinfo{author}{\bibfnamefont{H.-Y.} \bibnamefont{Jin}},
  \bibinfo{author}{\bibfnamefont{T.~G.} \bibnamefont{Steele}},
  \bibnamefont{and} \bibinfo{author}{\bibfnamefont{Z.-F.} \bibnamefont{Zhang}},
  \bibinfo{journal}{Phys. Rev.} \textbf{\bibinfo{volume}{D94}},
  \bibinfo{pages}{054037} (\bibinfo{year}{2016}),
  \eprint{arXiv:1608.03028[hep-ph]}.

\bibitem[{\citenamefont{Chetyrkin and Narison}(2000)}]{Chetyrkin:2000tj}
\bibinfo{author}{\bibfnamefont{K.~G.} \bibnamefont{Chetyrkin}}
  \bibnamefont{and} \bibinfo{author}{\bibfnamefont{S.}~\bibnamefont{Narison}},
  \bibinfo{journal}{Phys. Lett.} \textbf{\bibinfo{volume}{B485}},
  \bibinfo{pages}{145} (\bibinfo{year}{2000}), \eprint{hep-ph/0003151}.

\bibitem[{\citenamefont{Tang and Qiao}(2016)}]{Qiao:2015iea}
\bibinfo{author}{\bibfnamefont{L.}~\bibnamefont{Tang}} \bibnamefont{and}
  \bibinfo{author}{\bibfnamefont{C.-F.} \bibnamefont{Qiao}},
  \bibinfo{journal}{Nucl. Phys.} \textbf{\bibinfo{volume}{B904}},
  \bibinfo{pages}{282} (\bibinfo{year}{2016}),
  \eprint{arXiv:1509.00305[hep-ph]}.

\bibitem[{\citenamefont{Pimikov et~al.}(2017)\citenamefont{Pimikov, Lee,
  Kochelev, Zhang, and Khandramai}}]{Pimikov:2017bkk}
\bibinfo{author}{\bibfnamefont{A.}~\bibnamefont{Pimikov}},
  \bibinfo{author}{\bibfnamefont{H.-J.} \bibnamefont{Lee}},
  \bibinfo{author}{\bibfnamefont{N.}~\bibnamefont{Kochelev}},
  \bibinfo{author}{\bibfnamefont{P.}~\bibnamefont{Zhang}}, \bibnamefont{and}
  \bibinfo{author}{\bibfnamefont{V.}~\bibnamefont{Khandramai}},
  \bibinfo{journal}{Phys. Rev.} \textbf{\bibinfo{volume}{D96}},
  \bibinfo{pages}{114024} (\bibinfo{year}{2017}),
  \eprint{arXiv:1708.07675[hep-ph]}.

\bibitem[{\citenamefont{Jaffe}(1977{\natexlab{a}})}]{1977-Jaffe-p267-267}
\bibinfo{author}{\bibfnamefont{R.~L.} \bibnamefont{Jaffe}},
  \bibinfo{journal}{Phys. Rev.} \textbf{\bibinfo{volume}{D15}},
  \bibinfo{pages}{267} (\bibinfo{year}{1977}{\natexlab{a}}).

\bibitem[{\citenamefont{Jaffe}(1977{\natexlab{b}})}]{1977-Jaffe-p281-281}
\bibinfo{author}{\bibfnamefont{R.~L.} \bibnamefont{Jaffe}},
  \bibinfo{journal}{Phys. Rev.} \textbf{\bibinfo{volume}{D15}},
  \bibinfo{pages}{281} (\bibinfo{year}{1977}{\natexlab{b}}).

\bibitem[{\citenamefont{Maiani et~al.}(2004)\citenamefont{Maiani, Piccinini,
  Polosa, and Riquer}}]{2004-Maiani-p212002-212002}
\bibinfo{author}{\bibfnamefont{L.}~\bibnamefont{Maiani}},
  \bibinfo{author}{\bibfnamefont{F.}~\bibnamefont{Piccinini}},
  \bibinfo{author}{\bibfnamefont{A.}~\bibnamefont{Polosa}}, \bibnamefont{and}
  \bibinfo{author}{\bibfnamefont{V.}~\bibnamefont{Riquer}},
  \bibinfo{journal}{Phys.Rev.Lett.} \textbf{\bibinfo{volume}{93}},
  \bibinfo{pages}{212002} (\bibinfo{year}{2004}), \eprint{hep-ph/0407017}.

\bibitem[{\citenamefont{Chen et~al.}(2007{\natexlab{a}})\citenamefont{Chen,
  Hosaka, and Zhu}}]{2007-Chen-p94025-94025}
\bibinfo{author}{\bibfnamefont{H.-X.} \bibnamefont{Chen}},
  \bibinfo{author}{\bibfnamefont{A.}~\bibnamefont{Hosaka}}, \bibnamefont{and}
  \bibinfo{author}{\bibfnamefont{S.-L.} \bibnamefont{Zhu}},
  \bibinfo{journal}{Phys. Rev.} \textbf{\bibinfo{volume}{D76}},
  \bibinfo{pages}{094025} (\bibinfo{year}{2007}{\natexlab{a}}),
  \eprint{arXiv:0707.4586[hep-ph]}.

\bibitem[{\citenamefont{Chen et~al.}(2007{\natexlab{b}})\citenamefont{Chen,
  Hosaka, and Zhu}}]{2007-Chen-p369-372}
\bibinfo{author}{\bibfnamefont{H.-X.} \bibnamefont{Chen}},
  \bibinfo{author}{\bibfnamefont{A.}~\bibnamefont{Hosaka}}, \bibnamefont{and}
  \bibinfo{author}{\bibfnamefont{S.-L.} \bibnamefont{Zhu}},
  \bibinfo{journal}{Phys. Lett.} \textbf{\bibinfo{volume}{B650}},
  \bibinfo{pages}{369} (\bibinfo{year}{2007}{\natexlab{b}}),
  \eprint{hep-ph/0609163}.

\bibitem[{\citenamefont{Meyer and Swanson}(2015)}]{2015-Meyer-p21-58}
\bibinfo{author}{\bibfnamefont{C.~A.} \bibnamefont{Meyer}} \bibnamefont{and}
  \bibinfo{author}{\bibfnamefont{E.~S.} \bibnamefont{Swanson}},
  \bibinfo{journal}{Prog. Part. Nucl. Phys.} \textbf{\bibinfo{volume}{82}},
  \bibinfo{pages}{21} (\bibinfo{year}{2015}),
  \eprint{arXiv:1502.07276[hep-ph]}.

\bibitem[{\citenamefont{Thompson et~al.}(1997)}]{1997-Thompson-p1630-1633}
\bibinfo{author}{\bibfnamefont{D.}~\bibnamefont{Thompson}} \bibnamefont{et~al.}
  (\bibinfo{collaboration}{E852 Collaboration}),
  \bibinfo{journal}{Phys.Rev.Lett.} \textbf{\bibinfo{volume}{79}},
  \bibinfo{pages}{1630} (\bibinfo{year}{1997}), \eprint{hep-ex/9705011}.

\bibitem[{\citenamefont{Abele et~al.}(1999)}]{1999-Abele-p349-355}
\bibinfo{author}{\bibfnamefont{A.}~\bibnamefont{Abele}} \bibnamefont{et~al.}
  (\bibinfo{collaboration}{Crystal Barrel}), \bibinfo{journal}{Phys. Lett.}
  \textbf{\bibinfo{volume}{B446}}, \bibinfo{pages}{349} (\bibinfo{year}{1999}).

\bibitem[{\citenamefont{Adams et~al.}(2007)}]{2007-Adams-p27-31}
\bibinfo{author}{\bibfnamefont{G.~S.} \bibnamefont{Adams}} \bibnamefont{et~al.}
  (\bibinfo{collaboration}{E862}), \bibinfo{journal}{Phys. Lett.}
  \textbf{\bibinfo{volume}{B657}}, \bibinfo{pages}{27} (\bibinfo{year}{2007}),
  \eprint{hep-ex/0612062}.

\bibitem[{\citenamefont{Chen et~al.}(2008{\natexlab{a}})\citenamefont{Chen,
  Hosaka, and Zhu}}]{2008-Chen-p117502-117502}
\bibinfo{author}{\bibfnamefont{H.-X.} \bibnamefont{Chen}},
  \bibinfo{author}{\bibfnamefont{A.}~\bibnamefont{Hosaka}}, \bibnamefont{and}
  \bibinfo{author}{\bibfnamefont{S.-L.} \bibnamefont{Zhu}},
  \bibinfo{journal}{Phys.Rev.} \textbf{\bibinfo{volume}{D78}},
  \bibinfo{pages}{117502} (\bibinfo{year}{2008}{\natexlab{a}}),
  \eprint{arXiv:0808.2344[hep-ph]}.

\bibitem[{\citenamefont{Chen et~al.}(2008{\natexlab{b}})\citenamefont{Chen,
  Hosaka, and Zhu}}]{2008-Chen-p54017-54017}
\bibinfo{author}{\bibfnamefont{H.-X.} \bibnamefont{Chen}},
  \bibinfo{author}{\bibfnamefont{A.}~\bibnamefont{Hosaka}}, \bibnamefont{and}
  \bibinfo{author}{\bibfnamefont{S.-L.} \bibnamefont{Zhu}},
  \bibinfo{journal}{Phys. Rev.} \textbf{\bibinfo{volume}{D78}},
  \bibinfo{pages}{054017} (\bibinfo{year}{2008}{\natexlab{b}}),
  \eprint{arXiv:0806.1998[hep-ph]}.

\bibitem[{\citenamefont{Jiao et~al.}(2009)\citenamefont{Jiao, Chen, Chen, and
  Zhu}}]{2009-Jiao-p114034-114034}
\bibinfo{author}{\bibfnamefont{C.-K.} \bibnamefont{Jiao}},
  \bibinfo{author}{\bibfnamefont{W.}~\bibnamefont{Chen}},
  \bibinfo{author}{\bibfnamefont{H.-X.} \bibnamefont{Chen}}, \bibnamefont{and}
  \bibinfo{author}{\bibfnamefont{S.-L.} \bibnamefont{Zhu}},
  \bibinfo{journal}{Phys. Rev.} \textbf{\bibinfo{volume}{D79}},
  \bibinfo{pages}{114034} (\bibinfo{year}{2009}),
  \eprint{arXiv:0905.0774[hep-ph]}.

\bibitem[{\citenamefont{Huang et~al.}(2017)\citenamefont{Huang, Chen, Steele,
  Zhang, and Jin}}]{2017-Huang-p76017-76017}
\bibinfo{author}{\bibfnamefont{Z.-R.} \bibnamefont{Huang}},
  \bibinfo{author}{\bibfnamefont{W.}~\bibnamefont{Chen}},
  \bibinfo{author}{\bibfnamefont{T.~G.} \bibnamefont{Steele}},
  \bibinfo{author}{\bibfnamefont{Z.-F.} \bibnamefont{Zhang}}, \bibnamefont{and}
  \bibinfo{author}{\bibfnamefont{H.-Y.} \bibnamefont{Jin}},
  \bibinfo{journal}{Phys. Rev.} \textbf{\bibinfo{volume}{D95}},
  \bibinfo{pages}{076017} (\bibinfo{year}{2017}),
  \eprint{arXiv:1610.02081[hep-ph]}.

\bibitem[{\citenamefont{Du et~al.}(2013)\citenamefont{Du, Chen, Chen, and
  Zhu}}]{Du:2012pn}
\bibinfo{author}{\bibfnamefont{M.-L.} \bibnamefont{Du}},
  \bibinfo{author}{\bibfnamefont{W.}~\bibnamefont{Chen}},
  \bibinfo{author}{\bibfnamefont{X.-L.} \bibnamefont{Chen}}, \bibnamefont{and}
  \bibinfo{author}{\bibfnamefont{S.-L.} \bibnamefont{Zhu}},
  \bibinfo{journal}{Chin. Phys.} \textbf{\bibinfo{volume}{C37}},
  \bibinfo{pages}{033104} (\bibinfo{year}{2013}),
  \eprint{arXiv:1203.5199[hep-ph]}.

\bibitem[{\citenamefont{Shifman et~al.}(1979)\citenamefont{Shifman, Vainshtein,
  and Zakharov}}]{1979-Shifman-p385-447}
\bibinfo{author}{\bibfnamefont{M.~A.} \bibnamefont{Shifman}},
  \bibinfo{author}{\bibfnamefont{A.~I.} \bibnamefont{Vainshtein}},
  \bibnamefont{and} \bibinfo{author}{\bibfnamefont{V.~I.}
  \bibnamefont{Zakharov}}, \bibinfo{journal}{Nucl. Phys.}
  \textbf{\bibinfo{volume}{B147}}, \bibinfo{pages}{385} (\bibinfo{year}{1979}).

\bibitem[{\citenamefont{Reinders et~al.}(1985)\citenamefont{Reinders,
  Rubinstein, and Yazaki}}]{1985-Reinders-p1-1}
\bibinfo{author}{\bibfnamefont{L.~J.} \bibnamefont{Reinders}},
  \bibinfo{author}{\bibfnamefont{H.}~\bibnamefont{Rubinstein}},
  \bibnamefont{and} \bibinfo{author}{\bibfnamefont{S.}~\bibnamefont{Yazaki}},
  \bibinfo{journal}{Phys. Rept.} \textbf{\bibinfo{volume}{127}},
  \bibinfo{pages}{1} (\bibinfo{year}{1985}).

\bibitem[{\citenamefont{Colangelo and
  Khodjamirian}(2000)}]{2000-Colangelo-p1495-1576}
\bibinfo{author}{\bibfnamefont{P.}~\bibnamefont{Colangelo}} \bibnamefont{and}
  \bibinfo{author}{\bibfnamefont{A.}~\bibnamefont{Khodjamirian}},
  \bibinfo{journal}{Frontier of Particle Physics} \textbf{\bibinfo{volume}{3}}
  (\bibinfo{year}{2000}), \eprint{hep-ph/0010175}.

\bibitem[{\citenamefont{Narison}(1982)}]{Narison:1980ti}
\bibinfo{author}{\bibfnamefont{S.}~\bibnamefont{Narison}},
  \bibinfo{journal}{Phys. Rept.} \textbf{\bibinfo{volume}{84}},
  \bibinfo{pages}{263} (\bibinfo{year}{1982}).

\bibitem[{\citenamefont{Narison}(2002)}]{Narison:2002hk}
\bibinfo{author}{\bibfnamefont{S.}~\bibnamefont{Narison}}
  (\bibinfo{year}{2002}), \eprint{hep-ph/0202200}.

\bibitem[{\citenamefont{Shankar}(1977)}]{Shankar:1977ap}
\bibinfo{author}{\bibfnamefont{R.}~\bibnamefont{Shankar}},
  \bibinfo{journal}{Phys. Rev.} \textbf{\bibinfo{volume}{D15}},
  \bibinfo{pages}{755} (\bibinfo{year}{1977}).

\bibitem[{\citenamefont{Narison}(2009{\natexlab{a}})}]{Narison:2009vj}
\bibinfo{author}{\bibfnamefont{S.}~\bibnamefont{Narison}},
  \bibinfo{journal}{Phys. Lett.} \textbf{\bibinfo{volume}{B675}},
  \bibinfo{pages}{319} (\bibinfo{year}{2009}{\natexlab{a}}),
  \eprint{arXiv:0903.2266[hep-ph]}.

\bibitem[{\citenamefont{Matheus et~al.}(2007)\citenamefont{Matheus, Narison,
  Nielsen, and Richard}}]{Matheus:2006xi}
\bibinfo{author}{\bibfnamefont{R.~D.} \bibnamefont{Matheus}},
  \bibinfo{author}{\bibfnamefont{S.}~\bibnamefont{Narison}},
  \bibinfo{author}{\bibfnamefont{M.}~\bibnamefont{Nielsen}}, \bibnamefont{and}
  \bibinfo{author}{\bibfnamefont{J.~M.} \bibnamefont{Richard}},
  \bibinfo{journal}{Phys. Rev.} \textbf{\bibinfo{volume}{D75}},
  \bibinfo{pages}{014005} (\bibinfo{year}{2007}), \eprint{hep-ph/0608297}.

\bibitem[{\citenamefont{Lucha et~al.}(2018{\natexlab{a}})\citenamefont{Lucha,
  Melikhov, and Simula}}]{Lucha:2018ouu}
\bibinfo{author}{\bibfnamefont{W.}~\bibnamefont{Lucha}},
  \bibinfo{author}{\bibfnamefont{D.}~\bibnamefont{Melikhov}}, \bibnamefont{and}
  \bibinfo{author}{\bibfnamefont{S.}~\bibnamefont{Simula}}
  (\bibinfo{year}{2018}{\natexlab{a}}), \eprint{arXiv:1810.12569[hep-ph]}.

\bibitem[{\citenamefont{Lucha et~al.}(2018{\natexlab{b}})\citenamefont{Lucha,
  Melikhov, and Simula}}]{Lucha:2017zng}
\bibinfo{author}{\bibfnamefont{W.}~\bibnamefont{Lucha}},
  \bibinfo{author}{\bibfnamefont{D.}~\bibnamefont{Melikhov}}, \bibnamefont{and}
  \bibinfo{author}{\bibfnamefont{S.}~\bibnamefont{Simula}},
  \bibinfo{journal}{Eur. Phys. J.} \textbf{\bibinfo{volume}{C78}},
  \bibinfo{pages}{168} (\bibinfo{year}{2018}{\natexlab{b}}),
  \eprint{arXiv:1702.07537[hep-ph]}.

\bibitem[{\citenamefont{Shifman and Voloshin}(1985)}]{Shifman:1984wx}
\bibinfo{author}{\bibfnamefont{M.~A.} \bibnamefont{Shifman}} \bibnamefont{and}
  \bibinfo{author}{\bibfnamefont{M.~B.} \bibnamefont{Voloshin}},
  \bibinfo{journal}{Sov. J. Nucl. Phys.} \textbf{\bibinfo{volume}{41}},
  \bibinfo{pages}{120} (\bibinfo{year}{1985}), \bibinfo{note}{[Yad.
  Fiz.41,187(1985)]}.

\bibitem[{\citenamefont{Launer et~al.}(1984)\citenamefont{Launer, Narison, and
  Tarrach}}]{Launer:1983ib}
\bibinfo{author}{\bibfnamefont{G.}~\bibnamefont{Launer}},
  \bibinfo{author}{\bibfnamefont{S.}~\bibnamefont{Narison}}, \bibnamefont{and}
  \bibinfo{author}{\bibfnamefont{R.}~\bibnamefont{Tarrach}},
  \bibinfo{journal}{Z. Phys.} \textbf{\bibinfo{volume}{C26}},
  \bibinfo{pages}{433} (\bibinfo{year}{1984}).

\bibitem[{\citenamefont{Narison}(1995)}]{Narison:1995jr}
\bibinfo{author}{\bibfnamefont{S.}~\bibnamefont{Narison}},
  \bibinfo{journal}{Phys. Lett.} \textbf{\bibinfo{volume}{B361}},
  \bibinfo{pages}{121} (\bibinfo{year}{1995}), \eprint{hep-ph/9504334}.

\bibitem[{\citenamefont{Eidelman et~al.}(1979)\citenamefont{Eidelman, Kurdadze,
  and Vainshtein}}]{Eidelman:1978xy}
\bibinfo{author}{\bibfnamefont{S.~I.} \bibnamefont{Eidelman}},
  \bibinfo{author}{\bibfnamefont{L.~M.} \bibnamefont{Kurdadze}},
  \bibnamefont{and} \bibinfo{author}{\bibfnamefont{A.~I.}
  \bibnamefont{Vainshtein}}, \bibinfo{journal}{Phys. Lett.}
  \textbf{\bibinfo{volume}{B82}}, \bibinfo{pages}{278} (\bibinfo{year}{1979}).

\bibitem[{\citenamefont{Narison}(2012{\natexlab{a}})}]{Narison:2011xe}
\bibinfo{author}{\bibfnamefont{S.}~\bibnamefont{Narison}},
  \bibinfo{journal}{Phys. Lett.} \textbf{\bibinfo{volume}{B706}},
  \bibinfo{pages}{412} (\bibinfo{year}{2012}{\natexlab{a}}),
  \eprint{arXiv:1105.2922[hep-ph]}.

\bibitem[{\citenamefont{Narison}(2012{\natexlab{b}})}]{Narison:2011rn}
\bibinfo{author}{\bibfnamefont{S.}~\bibnamefont{Narison}},
  \bibinfo{journal}{Phys. Lett.} \textbf{\bibinfo{volume}{B707}},
  \bibinfo{pages}{259} (\bibinfo{year}{2012}{\natexlab{b}}),
  \eprint{arXiv:1105.5070[hep-ph]}.

\bibitem[{\citenamefont{Dosch et~al.}(1989)\citenamefont{Dosch, Jamin, and
  Narison}}]{Dosch:1988vv}
\bibinfo{author}{\bibfnamefont{H.~G.} \bibnamefont{Dosch}},
  \bibinfo{author}{\bibfnamefont{M.}~\bibnamefont{Jamin}}, \bibnamefont{and}
  \bibinfo{author}{\bibfnamefont{S.}~\bibnamefont{Narison}},
  \bibinfo{journal}{Phys. Lett.} \textbf{\bibinfo{volume}{B220}},
  \bibinfo{pages}{251} (\bibinfo{year}{1989}).

\bibitem[{\citenamefont{Ioffe}(1981)}]{Ioffe:1981kw}
\bibinfo{author}{\bibfnamefont{B.~L.} \bibnamefont{Ioffe}},
  \bibinfo{journal}{Nucl. Phys.} \textbf{\bibinfo{volume}{B188}},
  \bibinfo{pages}{317} (\bibinfo{year}{1981}), \bibinfo{note}{[Erratum: Nucl.
  Phys.B191,591(1981)]}.

\bibitem[{\citenamefont{Narison}(1993)}]{Narison:1992ru}
\bibinfo{author}{\bibfnamefont{S.}~\bibnamefont{Narison}},
  \bibinfo{journal}{Phys. Lett.} \textbf{\bibinfo{volume}{B300}},
  \bibinfo{pages}{293} (\bibinfo{year}{1993}).

\bibitem[{\citenamefont{Narison}(2009{\natexlab{b}})}]{Narison:2009vy}
\bibinfo{author}{\bibfnamefont{S.}~\bibnamefont{Narison}},
  \bibinfo{journal}{Phys. Lett.} \textbf{\bibinfo{volume}{B673}},
  \bibinfo{pages}{30} (\bibinfo{year}{2009}{\natexlab{b}}),
  \eprint{arXiv:0901.3823[hep-ph]}.

\bibitem[{\citenamefont{Narison}(2015)}]{NARISON2015189}
\bibinfo{author}{\bibfnamefont{S.}~\bibnamefont{Narison}},
  \bibinfo{journal}{Nuclear and Particle Physics Proceedings}
  \textbf{\bibinfo{volume}{258-259}}, \bibinfo{pages}{189 }
  (\bibinfo{year}{2015}), ISSN \bibinfo{issn}{2405-6014}, \bibinfo{note}{17th
  International Conference in Quantum Chromodynamics (QCD 14)}.

\end{thebibliography}

\end{document}